# RUSSIAN ACADEMY OF SCIENCES
## NATIONAL GEOPHYSICAL COMMITTEE

# РОССИЙСКАЯ АКАДЕМИЯ НАУК
## НАЦИОНАЛЬНЫЙ ГЕОФИЗИЧЕСКИЙ КОМИТЕТ

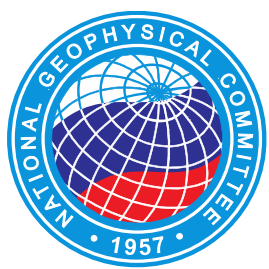

# NATIONAL REPORT

for the
International Association of Geodesy
of the
International Union of Geodesy and Geophysics
2019–2022

# НАЦИОНАЛЬНЫЙ ДОКЛАД

для
Международной ассоциации геодезии
Международного
геодезического и геофизического союза
2019–2022

**Москва   2023   Moscow**

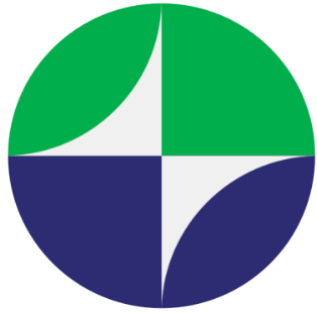

**Presented to the XXVIII General Assembly
of the
International Union of Geodesy and Geophysics**

**К XXVIII Генеральной ассамблее
Международного геодезического и геофизического
союза**

**RUSSIAN ACADEMY OF SCIENCES**

National Geophysical Committee

# NATIONAL REPORT

for the

International Association of Geodesy

of the

International Union of Geodesy and Geophysics

2019–2022

Presented to the XXVIII General Assembly

of the

IUGG

2023

Moscow

Major results of researches conducted by Russian geodesists in 2019–2022 on the topics of the International Association of Geodesy (IAG) of the International Union of Geodesy and Geophysics (IUGG) are presented in this issue. This report is prepared by the Section of Geodesy of the National Geophysical Committee of Russia. In the report prepared for the XXVII General Assembly of IUGG (Germany, Berlin, 11–20 July 2023), the results of principal researches in geodesy, geodynamics, gravimetry, in the studies of geodetic reference frame creation and development, Earth's shape and gravity field, Earth's rotation, geodetic theory, its application and some other directions are briefly described. For some objective reasons not all results obtained by Russian scientists on the field of geodesy are included in the report.

The following institutes participated in the preparation of the Report:

[1]Far Eastern Federal University, Vladivostok, Russia
[2]Federal Scientific-Technical Center of Geodesy, Cartography and Spatial Data Infrastructure, Moscow, Russia
[3]Geophysical Center of the Russian Academy of Sciences, Moscow, Russia
[4]Institute for Applied Mathematics, Vladivostok, Russia
[5]Joint Stock Company «Central Research Institute for Machine Building», Moscow Region, Russia
[6]Moscow State University of Geodesy and Cartography, Moscow, Russia
[7]National Research Institute for Physical-Technical and Radio Engineering Measurements (VNIIFTRI), Mendeleevo, Moscow Reg., Russia
[8]Peoples' Friendship University of Russia named after Patrice Lumumba, Moscow, Russia
[9]Pulkovo Observatory, Saint Petersburg, Russia
[10]Russian society of geodesy, cartography and land management, Moscow, Russia
[11]Schmidt Institute of Physics of the Earth of the Russian Academy of Sciences Moscow, Russia
[12]Sternberg Astronomical Institute, Lomonosov Moscow State University, Moscow, Russia

The Report was prepared by the following scientists:

[9]Gorshkov V., [5]Gusev I., [8]Dokukin P., [3]Kaftan V., [9]Malkin Z., [2]Mazurova E., [11]Mikhailov V., [7]Pasynok S., [10]Pobedinsky G., [2]Popadyev V., [6]Savinykh V., [1,4]Shestakov N., [2]Stoliarov I., [6]Sugaipova L., [12]Zotov L.

В данном Национальном докладе представлены основные результаты исследований, проводимых российскими геодезистами в 2019—2022 гг., по темам, соответствующим направлениям деятельности, координируемым Международной ассоциацией геодезии (МАГ) Международного геодезического и геофизического союза (МГГС). Данный доклад подготовлен Секцией геодезии Национального геофизического комитета Российской академии наук. В данном докладе, подготовленном к XXVIII Генеральной ассамблее МГГС (Германия, Берлин, 11—20 июля 2023 г.), представлены основные результаты исследований в области геодезии, геодинамики, гравиметрии, создания геодезических систем отсчета, формы и гравитационного поля Земли, вращения Земли, теории геодезии и ее приложений. По понятным причинам, в отчет были включены не все результаты, полученные российскими учеными в области геодезии.

В подготовке данного отчета принимали участие представители следующих организаций:

[1]Дальневосточный федеральный университет, Владивосток, Россия
[2] Федеральный научно-технический центр геодезии, картографии и инфраструктуры пространственных данных, Москва, Россия
[3]Геофизический центр РАН, Москва, Россия
[4]Институт прикладной математики РАН, Владивосток, Россия
[5]Акционерное общество «Центральный научно-исследовательский институт машиностроения» (АО «ЦНИИмаш»), Королев, Московская обл., Россия
[6]Московский государственный университет геодезии и картографии, Москва, Россия
[7]Всероссийский научно-исследовательский институт физико-технических и радиотехнических измерений (ВНИИФТРИ), Менделеево, Московская обл., Россия
[8]Российский университет дружбы народов им. Патриса Лумумбы, Москва, Россия
[9]Главная (Пулковская) астрономическая обсерватория РАН, Санкт-Петербург, Россия
[10]Российское общество геодезии, картографии и землеустройства, Москва, Россия
[11] Институт физики Земли им. О.Ю. Шмидта  Российской академии наук, Москва, Россия
[12]Государственный астрономический институт им. П.К. Штернберга  Московского государственного университета, Москва, Россия

В подготовке данного отчета принимали участие:

[9]Горшков В., [5]Гусев И., [8]Докукин П., [3]Кафтан В., [9]Малкин З., [2]Мазурова Е., [11]Михайлов В., [7]Пасынок С., [10]Побединский Г., [2]Попадьев В., [6]Савиных В., [1,4]Шестаков Н., [2]Столяров И., [6]Сугаипова Л., [12]Зотов Л.

# From the Science Editors of the National Report on Geodesy

Savinykh V.[1], Kaftan V.[2]

[1]Moscow State University of Geodesy and Cartography, Moscow, Russia
[2]Geophysical Center of the Russian Academy of Sciences, Moscow, Russia

This National Report to the General Assembly of the IAG / IUGG contains information on the main achievements of Russian researchers in the field of geodesy for the period 2019–2022. The information is collected from published works and is a collection of short abstracts of selected scientific publications. The content of this analytical compendium briefly presents the results of major research and development in the fields of geodesy, geodynamics, gravimetry, the creation of geodetic reference frames, the study of the figure and gravitational field of the Earth, the parameters of the Earth's rotation, the development of geodetic theories, geodetic applications and some related areas of research. The information is presented in accordance with the main topics of scientific research, summarized by the International Association of Geodesy of the International Union of Geodesy and Geophysics. The book contains information about the scientific work of not only Russian researchers, but also research on international projects with their participation.

The authors of the publication are members of the Geodesy Section of the National Geophysical Committee of the Russian Academy of Sciences. Of course, the content of the publication does not cover the entire volume of Russian research in the field of geodesy. Nevertheless, the authors hope that the most fundamental topics are reflected in the issue.

The most significant scientific and scientific and technical results include the following research and development.

The development of the national coordinate, height and gravimetric frames of the Russian Federation is the most important scientific and technical measure for the distribution of the geocentric coordinate system, the system of normal heights and the gravimetric system on the territory of the state. The adjustment of the Fundamental astro-geodesic network (FAGN) for 2014–2022 in the ITRF2014 system was carried out and inter-annual and intra-annual parameters of coordinate changes were derived.

On the territory of the Russian Federation, laser systems of a new generation are installed, capable of providing millimeter accuracy in measuring slope distances.

In addition to the tasks of determining the parameters of the Earth's rotation, the system provides the possibility of laser transmission of time, and the determination of differences in onboard time scales. The introduction of the presented measuring tools and methods, as well as proven methods for calibrating and verifying satellite laser ranging stations, allows you to rise to a qualitatively new level.

Very interesting are the results of the analysis of the relationship between changes in the parameters of the Earth's rotation and strong earthquakes. Studies of the correlation between the rotation and the Earth's magnetic field also deserve special attention. The key process in these interactions is the free nutation of the Earth's core.

Extensive original research has been done on the evolution of the Chandler Oscillation over the course of a century. The interrelation of the movement of the pole and the length of day (LOD) with geophysical processes, in particular with climatic changes, has been studied.

New possibilities of satellite altimetry have been explored. In particular, the economic advantages of determining the characteristics of the gravitational field in the waters of the World Ocean using satellite altimetry are shown. A method for determining the gravitational field in inaccessible areas based on airborne gravimetry data is proposed.

Determination of geoid heights from gravity field disturbances according to regional gravimetric measurements was carried out using a modified Hotine-Koch transformation. The numerical experiment demonstrated the high accuracy of the interpolation.

The theory of Molodensky is improved in the problem of solving the task of determining the anomalous geopotential on the basis of mixed anomalies of the gravitational field. New evidence has been obtained of the advantage of normal heights over orthometric ones in the territories of large states.

A number of interesting results have been obtained in solving problems of geodynamics by geodetic methods.

The use of long-term time series of continuous determinations of the coordinates of GNSS stations made it possible to study the evolution of movements and deformations of the earth's crust in seismically active regions. This led to the emergence of a new direction in the study of kinematic and dynamic processes in the upper parts of the earth's crust, by analogy with synoptic hydrometeorological studies. At the same time, slow deformation waves of total shear deformation and accumulation of a deficit of internal movements before strong earthquakes were registered. In particular, it should be noted that the area of small internal movements was discovered by Russian-Turkish researchers after the fact in the area of the devastating Turkish-Syrian earthquakes in February 2023.

According to continuous GNSS measurements in regional networks, international Russian-Chinese studies of the kinematics of the earth's crust in the Far East region and China are being carried out. The high efficiency of the use of continuous GNSS measurements in regional networks to improve the reliability of tsunami early warning has been confirmed. At the same time, it was proved that taking into account the layering and sphericity of the Earth model provides a more accurate estimate of the tsunami heights.

Various stages of the seismic cycle have been studied using data from complex observations by means of GNSS and the GRACE mission. Post-seismic and co-seismic processes are studied according to the gravitational field data in time for a number of mega-earthquakes in subduction zones..

SAR interferometry data were used to study the vertical movements of the earth's crust in the areas of active volcanoes in Kamchatka, as well as above underground workings. The rate of solidification of pyroclastic lava flows after volcanic eruptions has been studied. With a high degree of detail, areas of landslides and subsidence of soils in the coastal areas of Greater Sochi were identified.

In the tasks of the integrated use of new generation VLBI, GNSS and SLR, new algorithms for detecting gross measurement errors at the preprocessing level were applied in order to achieve millimeter accuracy.

On the territory of Eastern and Northern Europe, the movements and deformations of the earth's crust are studied according to the data of the regional integrated GNSS network using the created database on the speeds of movements.

On the territory of the Far East, studies are continuing on the concentration of water vapor in the surface layer of the atmosphere according to GNSS data and their relationship with atmospheric phenomena, forest fires, industrial pollution, etc. The dynamics of water vapor concentration according to GNSS data is being studied in the North-West region of Russia.

The effects of ionospheric-lithospheric relations have been studied using continuous GNSS data in the Russian Far East and foreign territories. The connection of ionospheric disturbances with a volcanic eruption and an underground nuclear test in South Korea has been studied.

***Section of geodesy of the National Geophysical Committee***
*Dr. V.P. Savinykh, Chairman, Academician of RAS*
*Dr. V.I. Kaftan, Vice-chairman*

# Reference Frames

Kaftan V.[1], Malkin Z.[2], Mazurova E.[3], Pasynok S.[4], Pobedinsky G.[5], Stoliarov I.[3]
Popadyev V. V.[3]

[1]Geophysical Center of the Russian Academy of Sciences, Moscow, Russia
[2]Pulkovo Observatory, Saint Petersburg, Russia
[3]Federal Scientific-Technical Center of Geodesy, Cartography and Spatial Data Infrastructure, Moscow, Russia
[4]National Research Institute for Physical-Technical and Radio Engineering Measurements (VNIIFTRI), Mendeleevo, Moscow Reg., Russia
[5]Russian society of geodesy, cartography and land management, Moscow, Russia

### *International Celestial Reference Frame*

Activities of the IAG Sub-commission 1.4: Interaction of Celestial and Terrestrial Reference Frames (Chair Zinovy Malkin) include investigations of various issues limiting the consistent realization of terrestrial and celestial reference frames at the mm/µas level of accuracy, such as:

• Insufficient number and non-optimal distribution of active and stable stations (VLBI and SLR in the first place) and radio sources.
• Technological (precision) limitations of existing techniques.
• Incompleteness of the theory and models.
• Not fully consistent models applied during data analysis.
• Not fully understood and agreed-upon details of the processing strategy.
• Not fully understood and accounted for the systematic errors of different techniques.

An overview of these activities is given in Malkin et al. [2019c], Malkin et al. [2020d].

A new method for subdivision of a spherical surface into equal area isolatitudinal grid was proposed in Malkin (2019a). The method is based on dividing a sphere into latitudinal rings of near-constant width with further splitting of each ring into equal-area cells. It is simple in construction and use and provides more uniform width of the latitudinal rings than other methods of similar pixelization a spherical surface. Examples of SREAG grids are shown in Fig. 1. The method can be equally useful for processing data sets given both on the celestial and terrestrial sphere. Some supplement features of this method were discussed in Malkin [2020c]. In Malkin [2021b], this method was used for selection of prospective sources for selection of new ICRF sources.

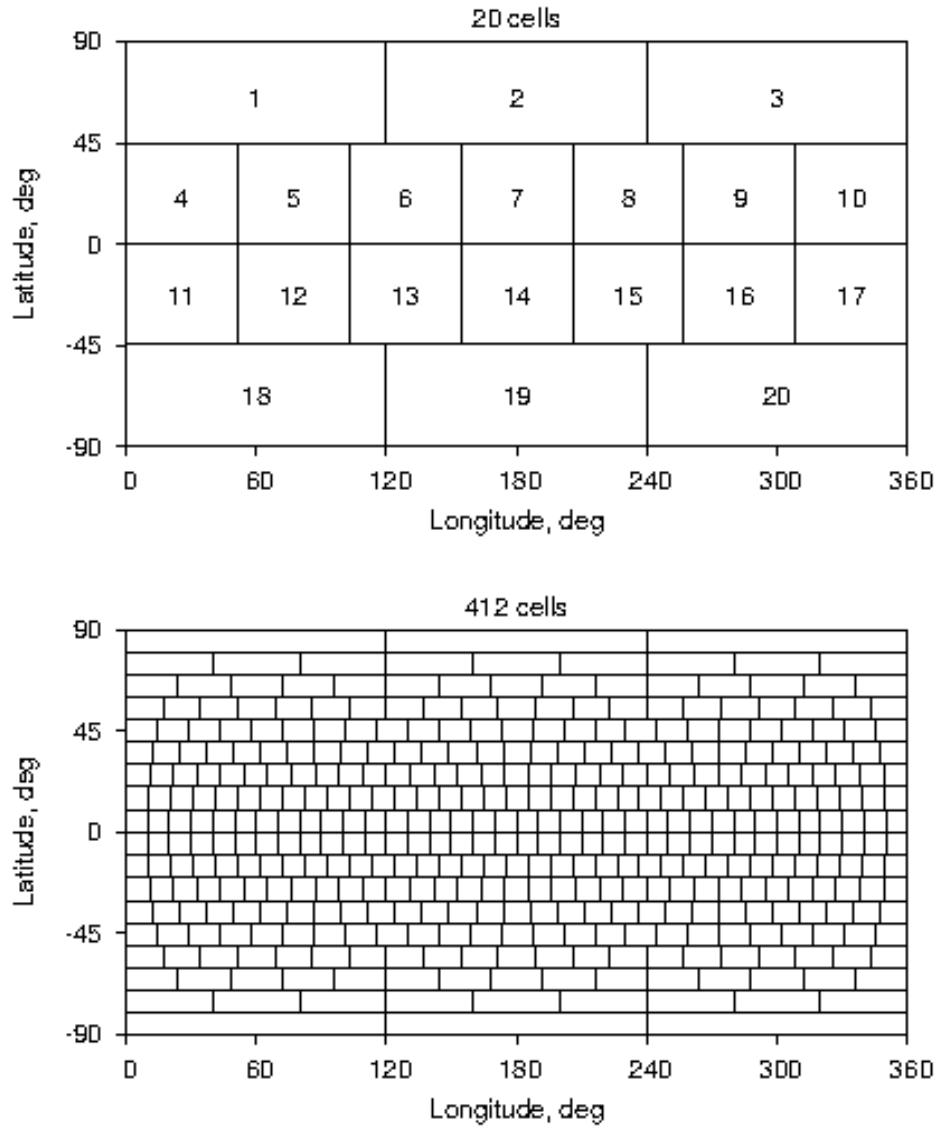

Fig. 1. Two examples of SREAG grids.

A new method of determination of orientation parameters between two spherical coordinate frames practically insensitive to outliers was proposed in Malkin [2021c]. This method is based on binning data over the equal-area cells of SREAG grid, followed by median filtering of the data in each cell. After this, a new data set is formed, consisting of data points near-uniformly distributed over the sphere. The vector spherical harmonics (VSH) decomposition is then applied to this data to finally compute the orientation parameters between two frames. The proposed method can be effectively used for comparison of both celestial and terrestrial frames. Fig. 2 shows an example of using this method for VLBI-based ICRF3 and space-based Gaia-CRF3 catalogs.

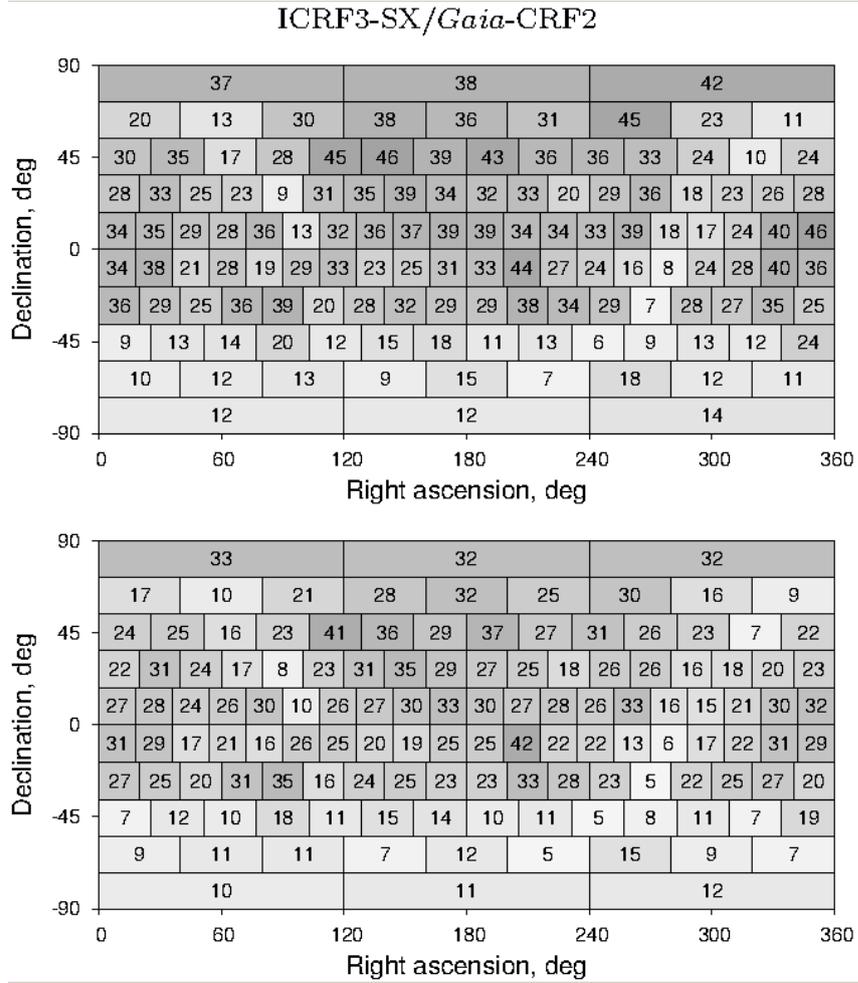

Fig. 2. Sky distribution of the sources in ICRF3 and Gaia-CRF3 catalogs.

The paper MacMillan et al. [2019b] is devoted to analysis of the impact of Galactic aberration on time evolution of the ICRF catalog of Galactic aberration in IVS analysis. It is caused by the circular motion of the solar system barycenter around the Galactic center and leads to a movement of the ICRF source up to about 5 µas/yr depending on the source position, which is quite substantial for the current accuracy of VLBI observations over decadal time scales. One of the specific goals of this work was to recommend a Galactic aberration model to be applied in the generation of ICRF3 as well as in other IVS analysis. After comparison of different approaches, the recommended value was proposed 5.8 µas/yr as estimated directly in a global VLBI along with ICRF3 solution.

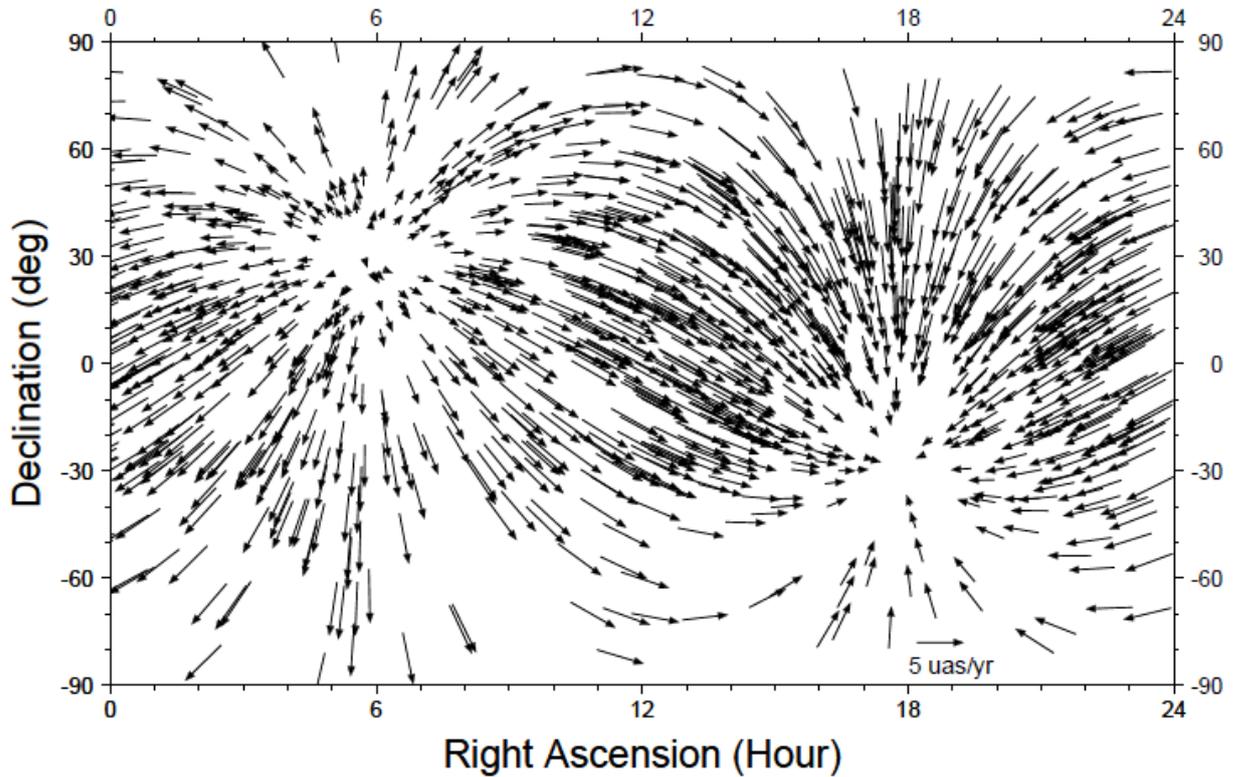

Fig. 3. Impact of the Galactic aberration on the source proper motion.

In 2019, the International VLBI Service for Geodesy and Astrometry (IVS) celebrated its 20-year jubilee. The paper Malkin [2020a] analyzed the dynamics of IVS development based on the statistical processing of the observational data collected in GAO RAN in 1979–2018. Statistics for observation years, stations, baselines, and radio sources were given. The evolution of observation statistics and the accuracy of the results obtained from the processing of VLBI observations is considered. As an example, Fig. 4 shows the time evolution of baseline length and station coordinates.

### *National Reference Frames*

Consideration of national coordinate systems, heights, gravimetric measurements and national geodetic, leveling and gravimetric networks are closely related to the concept of a geodetic support system. The concept of "System of geodetic support" has evolved with the development of geodesy, as well as any area of scientific and practical human activity, under the influence of two main factors: the demand in society at this stage of economic development and the level of technical means for the implementation of this activity.

The work [Brovar et al., 2022] highlights the fundamental basis for creating a highly efficient system of geodetic support and tactical problems of geodesy. The tactical tasks of geodesy follow from its main scientific task – determining the shape and size of the Earth, its gravitational field and their changes in time. Gravimetry and space geodesy provide data on the external gravitational field in time, astronomy – on the high-precision orientation of the coordinate system in time, geodesy (higher and lower) – on the physical surface and elements of the Earth's gravitational field in time, geodynamics – on changes in time of coordinates of points on the earth surface and characteristics of the Earth's gravitational field, metrology – about reference, samples and working means of geodetic (in the broad sense) measurements, thereby ensuring the unity of measurements.

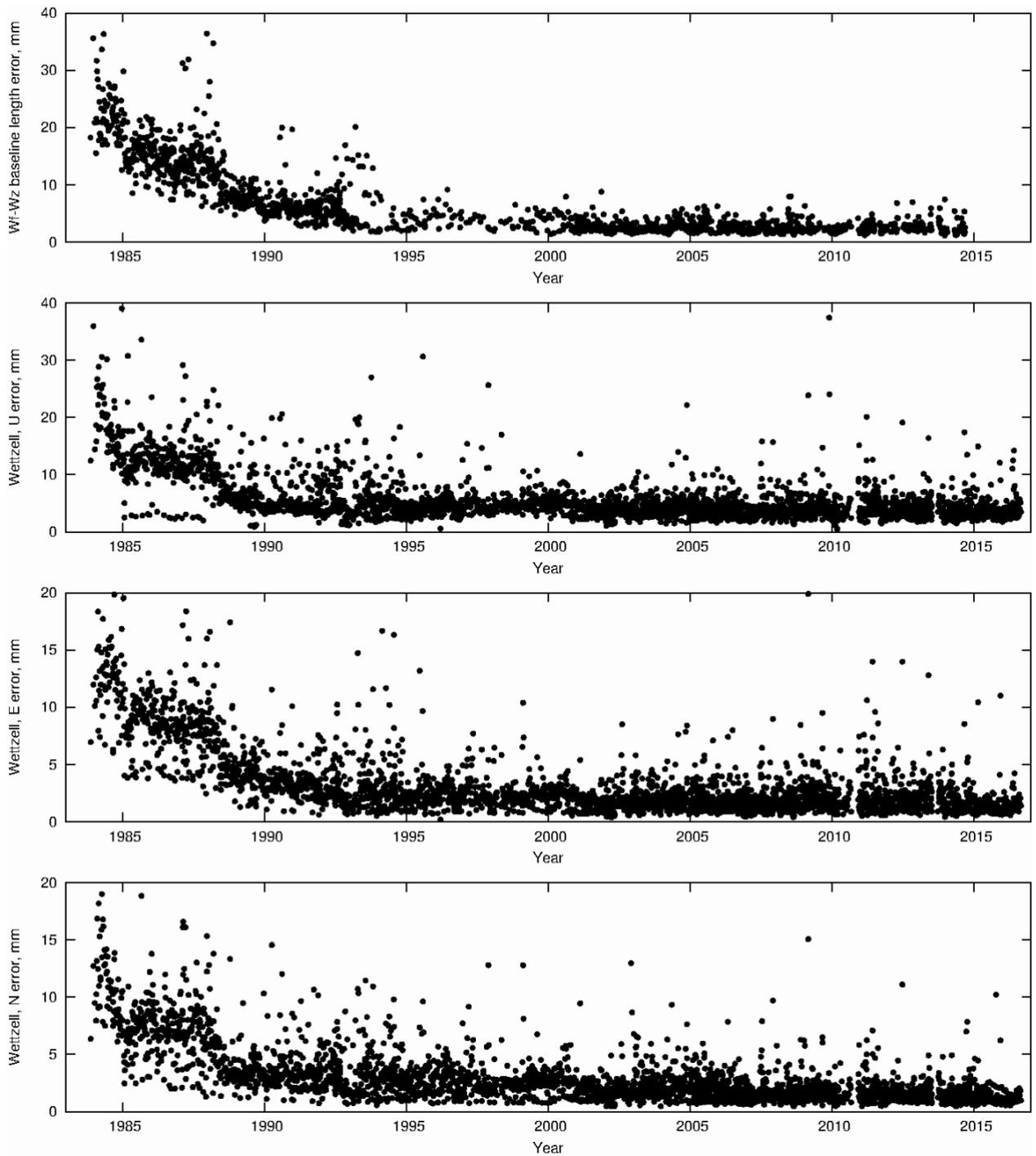

Fig. 4. Time evolution of the Westford–Wettzell baseline (upper plot) and UEN Wettzell coordinates.

In the general case, the geodetic support of the Russian Federation includes a unified state system of geodetic support; geodetic support of departmental, regional and municipal significance, including geodetic support of cadastral works; geodetic support for special purposes, which includes geodetic thickening networks created to solve the problems of engineering and geodetic surveys, the development of urban engineering and architectural infrastructure, to provide observations of displacements of buildings and structures, survey of underground utilities, geodetic work during the construction and operation of industrial facilities.

The unified state system of geodetic support includes:

- state coordinate, altitude and gravimetric bases;

- a system for providing consumers with the information necessary to accurately determine the location of objects in near real time;

- a system for determining the parameters of the figure of the Earth and the external gravitational field;

- a system for monitoring deformations of the earth's surface.

The evolution of the concept of "System of geodetic support" is considered in [Pobedinsky et al., 2022a], which outlines the basic principles for constructing a modern system of geodetic support:

- the full potential of modern measuring instruments should be used - satellite and ground-based geodetic, leveling, gravimetric and astronomical, based on different physical principles;

- the system of geodetic support should be maximally focused on the use of modern information and telecommunication technologies.

The work [Pobedinsky et al., 2022a] presents the general preliminary structure of the network information technology system for geodetic support of the Russian Federation and notes that certain elements of this structure are already functioning on the website of the Center for Geodesy, Cartography and SDI. This is information about the location of the points of the Fundamental Astronomical and Geodetic Network (FAGN), the values of the coordinates of the working centers of the FAGN points and the rate of their change in the GSK-2011 coordinate system for the epoch of January 1, 2011 https://rgs-centre.ru/fags-coords, list of points of the High Precision Geodetic Network (PGN) https://cgkipd.ru/opendata/spisok-punktov-vysokotochnoy-geodezicheskoy-seti-vgs/, Global Model of the Earth's Gravitational Field GAO2012 https://cgkipd.ru/opendata/gao2012/, metadata of reports on the creation of geodetic networks for special purposes https://cgkipd.ru/opendata/GSSN/ and information about operators of private differential geodetic stations included in the federal network of differential geodetic stations (FNGS) https://cgkipd.ru/opendata/ apk-fsgs.php;

The main part of the geodetic support system is the system of coordinates, heights, gravimetric measurements and, accordingly, geodetic, leveling and gravimetric networks.

A review and analysis of the state of geodetic coordinate systems of different levels, from global to regional and local, is given in [Pobedinsky, Kaftan, 2020].

Historical information is presented and a description of modern geodetic coordinate systems and their implementations is given. Proposals for the creation of a new special subcommittee of the International Association of Geodesy are highlighted. The features of the relationship of coordinate systems of different levels are revealed and recommendations are given for improving coordinate systems for state purposes. A list of problems and tasks facing the state, contributing to the harmonization and improvement of coordinate systems in the aspect of state geodetic support of the Russian Federation, is presented.

In [Kaftan et al., 2022], three implementations of the global (general-terrestrial) coordinate systems (reference) ITRF, WGS-84, and PZ-90 are considered. It is shown that highly developed countries with a large territory create national satellite geodetic networks that implement national coordinate systems. The rationale for the need to maintain and develop the national state geodetic coordinate system GSK-2011 for geodetic and cartographic work on the territory of the country, with the simultaneous participation of the Russian Federation in the creation of a global geodetic coordinate system (reference) at the UN initiative, is given.

Proposals for the participation of the Russian Federation in the creation of a unified global geodetic coordinate system are presented in [Pobedinsky, 2019, Pobedinsky et al., 2020b, c]. Conceptual concepts are considered and analyzed. Differences in terms and interpretations are presented. The state management decisions on state geodetic support are analyzed and proposed, and its place in the creation of a global coordinate system is determined.

The state and prospects for the development of the State primary standard of time units, frequency and the national time scale are considered in [Denisenko et al., 2021]. The problems of metrological support for GNSS geodetic equipment in the field of measuring large distances are considered in [Pobedinsky et al., 2021]. Proposals for the organization of metrological polygons on the basis of permanent FAGN points are presented. The inclusion of baselines several hundred kilometers long in such polygons opens up the possibility of estimating for certified receivers the error values proportional to the length of the measured lines at the level of the first units of $10^{-8}$. At the same time, the presence of short baselines (several tens or hundreds of meters long) in the composition of metrological polygons makes it possible to estimate satellite measurement errors that do not depend on the length of the determined lines.

The system of state geodetic reference frame of Russia can be divided into components:

- The state geodetic reference frame;
- The state leveling reference frame of I and II classes;
- The state gravimetric reference frame.

The state geodetic reference frame consist of a satellite geodetic reference frame includes the Fundamental Astronomical-Geodetic Network (FAGN), High-Precision Geodetic Network (PGN), the first-class Satellite Geodetic Network (SGN-1) as well as the existing ground-based geodetic networks of the first and second classes are used for densification of the FAGN, PGN and SGN-1.

Satellite geodetic reference frame has three levels:

The first level is materialized by the fundamental astronomical-geodetic network (FAGN), which has the highest priority in providing the navigational support and dissemination of coordinates in Russia. It is used for solving various scientific and technological problems which require precise navigation and timing. The FAGN is the primary geodetic basis for future improvement of the accuracy of the station coordinates of the National Reference Frame. Nowadays the FAGN consists of 99 permanent GNSS points (Fig. 5). Note that in Fig. 5, the double triangles denote a point having two working centers. Fig. 6 shows such a points, of which 98 are located on the territory of the Russian Federation and 1 point in Barentsburg at the Spitsbergen archipelago (Urban East Norwegian). The coordinates of this network are permanently being determined. The velocity and direction of movement of points are calculated at the end of the year. The average distance between the adjacent points of the FAGN is from 150 to 650 km. The normal heights are measured by the leveling of the second class with the rms error about 2.0 mm per 1 km. The gravity acceleration is measured with the most precise gravimeters in compliance with the requirements of the fundamental gravimetric survey. The values of the normal heights and the absolute acceleration of gravity are checked at each reference point every 5 or 8 years.

The second level is materialized by the High-Precision Geodetic Reference Network (PGN). The operational mission of this reference frame is to disseminate the ITRF to the entire territory of Russia and to refine the parameters of the relative orientation of the axes of the celestial and geodetic coordinate systems. Three hundred and eighty-eight points of PGN form an accurate homogeneous geodetic spatial network with an average distance between the adjacent points from 150 to 300 km.

Coordinates of the PGN points are determined by the space geodesy techniques that warrant the accurate measurement of the relative distances between points with the rms error not exceeding 3 mm $\pm$ 0.5D for each horizontal coordinate, and 5 mm $\pm$ 0.7D for a vertical coordinate - the geodetic height, where D is the distance between the points in km. Each point of the PGN is linked through various types of measurements with a number of other PGN points and with, at least, three of the closest points of the FAGN. In remote regions, where the points of the PGN are sparsely distributed and scarcely connected, it is required to link the PGN points with a larger number of the FAGN points and to monitor their mutual collocation more frequently.

Each point of the PGN is labeled with a normal height and the absolute value of the acceleration of gravity. The normal heights of the PGN points are given with respect to the benchmarks of the first or second-class leveling network or superimposed on the corresponding leveling lines.

The number of the existing points of the PGN reference frame gradually increases as the work on the densification of the frame goes on. In order to maintain the current accuracy of the expanding network, the relative positions of the new reference points are measured with the precision not less than 2 cm in each coordinate.

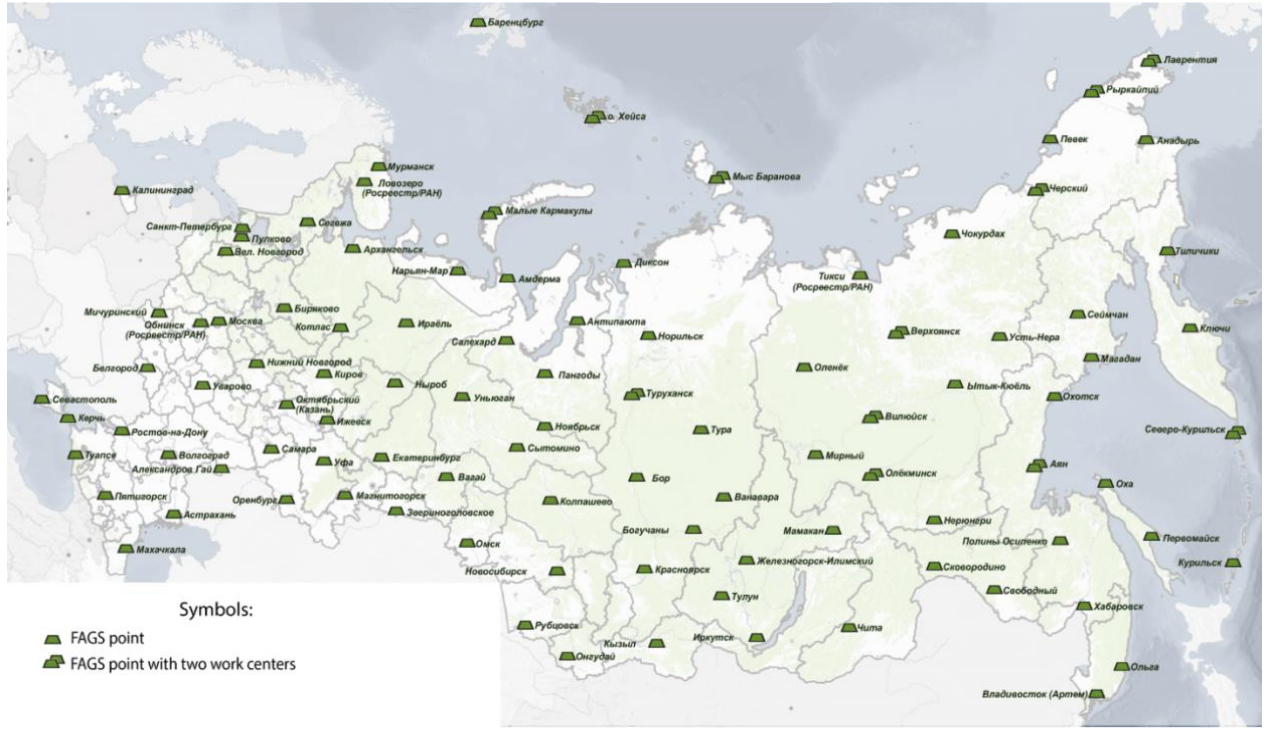

Fig. 5. FAGN stations as of 01.01.2023

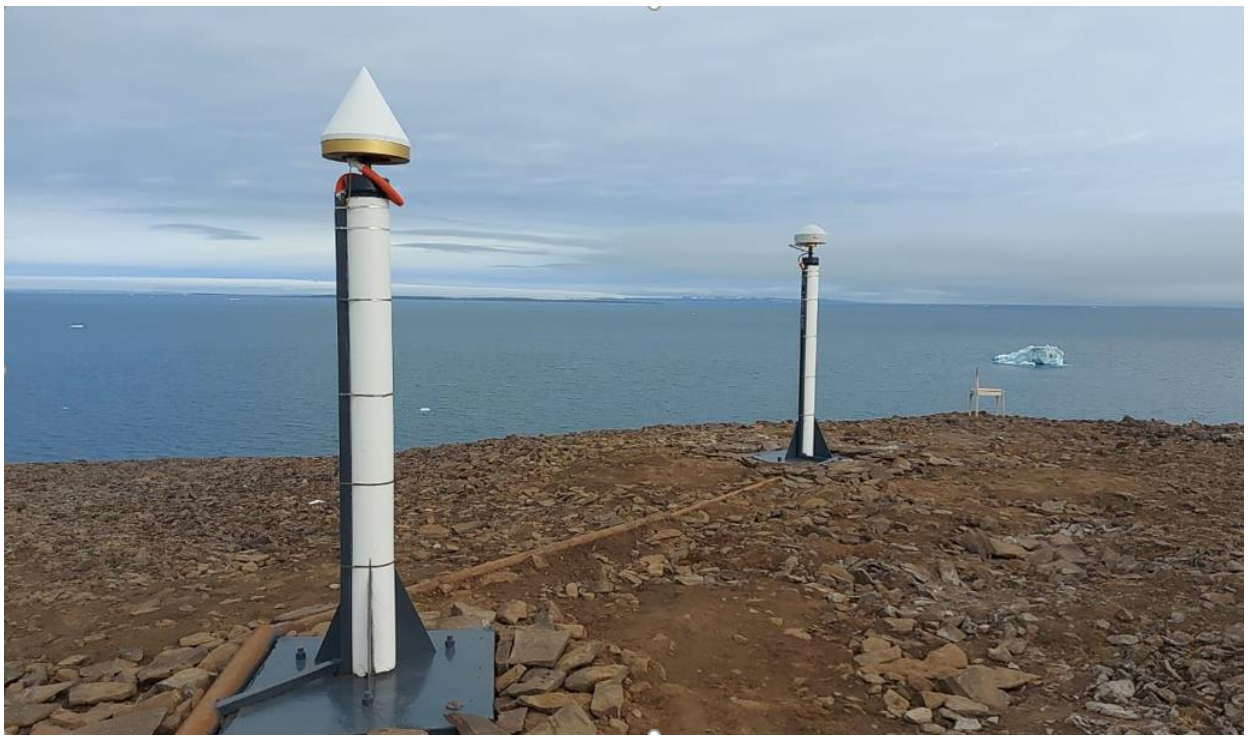

Fig. 6. FAGN station on the Hays Island of the Franz Josef Land archipelago with two work centers

The third level is the first-class satellite geodetic network (Satellite Geodetic Network of the 1st class SGN-1) that consists of 5 860 points with an average distance between the adjacent points from 25 to 35 km. The main purpose for introducing the third-level reference frame is to densify the first and second level networks, and to provide the most optimal conditions for utilization of the high precision and operational capabilities of the satellite positioning measurements for the ground-based geodetic applications.

The SGN-1 has been created by employing various techniques of space geodesy that are capable of measuring the relative positions of the adjacent points with the rms error of 3 mm ± 0.1D in the horizontal coordinates and 5 mm ± 0.2D in the geodesic height, where D is the distance between the points in km. The normal height of each point of the SGN-1 is determined either from the spirit leveling (with an accuracy being consistent with the requirements to the second or third-class leveling networks) or from the satellite-geodetic leveling as the difference between the geodetic and quasi-geoid heights.

The resulting accuracy of the point coordinates of the SGN-1 is established by the measurements conducted as many times as required by the state engineering service regulations. The link between the points is usually provided and maintained by the methods of space geodesy with the precision not less than 2 cm in the horizontal coordinates and 1 cm in the geodetic height. The average distance between the points of the FAGN and VGN, which coincide with or are linked to the points of the SGN-1 should not exceed 70 km in the populated regions of the country and 100 km in the uninhabited areas. The distance between the leveling benchmarks used for connection to the points of the SGN-1 is not greater than 100 km.

The points of the SGN-1 that are combined or linked with the benchmarks of the first, second and third-class leveling networks are used to refine the map of the quasi-geoid heights. As a rule, the points which normal heights have been determined only from the trigonometric leveling are not allowed to be used for determination of the heights of the SGN-1 points. Exceptions are applied to the points in the areas having no sufficient data on the quasi-geoid heights. In this case, the rms error in measuring the relative position of the adjacent points in the elevation should not exceed 20 cm.

The structure of the State Geodetic Reference Frame has a potential for further improvement by including new satellite navigation and geodetic stations which are built by the federal agencies and academic institutions in accordance with the Russian Government Regulations. Coordinates of the new stations are calculated and added to the points of the existing reference frame by the computer data processing centers being responsible for the maintenance of the federal differential satellite network.

*Geodetic reference frame constructed by classical terrestrial methods*

The State Geodetic Network of the 1–4 classes constructed by classical terrestrial methods: triangulation and polygonometry of classes 1–4. This network has retained its value, it is maintained in working condition and consists of about half of the million points.

*Federal Network of Deferential Geodetic Stations (FNGS)*

Russian Continuously Operating Geodetic Stations is a network of stations that provide global navigation satellite system data consisting of carrier phase and code range measurements in support of three dimensional positioning, meteorology, space weather, cadastral and geophysical applications throughout the Russia.

This network combines private and corporate stations into a single three-dimensional structure, the coordinates of which are determined by joint adjustment with the state FAGN network. At the same time, the FAGN is the basis of this general construction.

FNGS began to be implemented in 2021 and now comprises 2,079 stations (Fig. 7) of various organizations across the country and continues to expand. These stations are managed remotely. Each owner (participant) provides their data to a single center, which analyses it, processes it, and distributes it. FSGS data are used by surveyors, cadasters, geophysicists, meteorologists and other scientists.

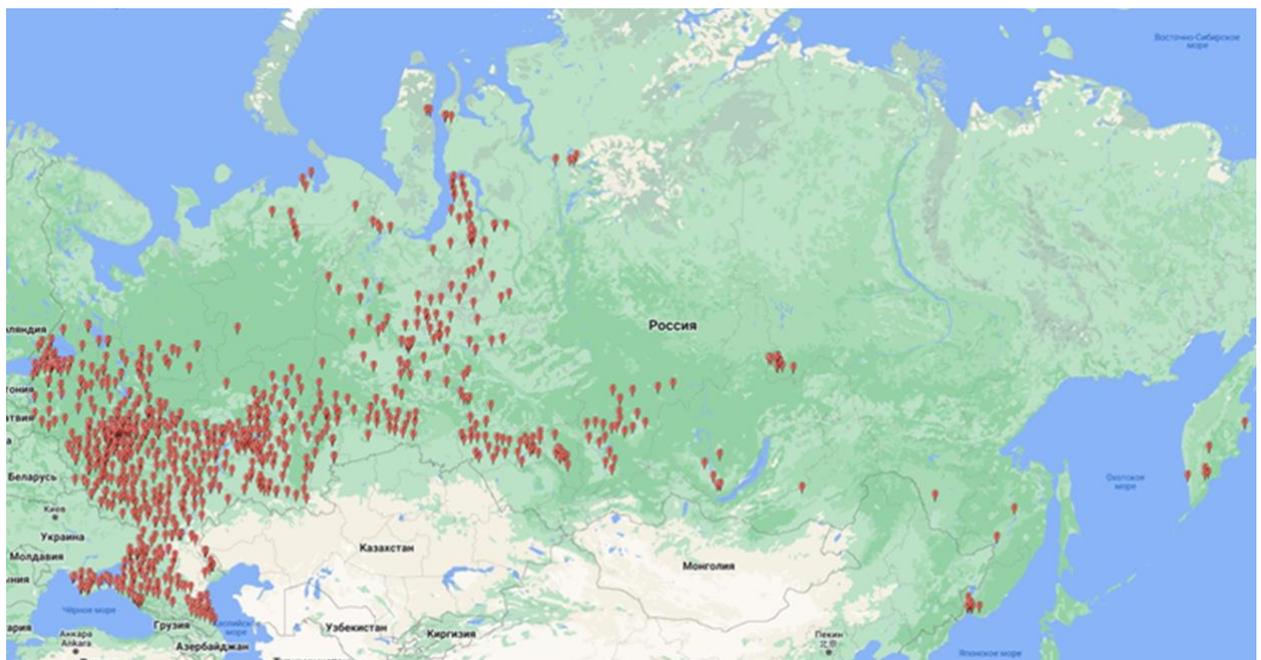

Fig. 7. Federal Network of Differential Geodetic Stations

The FNGS provides a single geospatial and timing basis for high-precision real-time positioning.

*The Main Russian Vertical Reference Frame (MRVRF) characteristics as of 01.05.2023*:

The MRVF consists of 105,000 benchmarks, of which 58,867 points of class I and 45,606 points of class II.

This network includes 188 polygons of class I, 880 polygons of class II and mixed polygons of classes I and II (see Fig. 8).

The average perimeter of a class I landfill for the territory of the Russian Federation is 1.5 thousand km. Moreover, in the European part of 1 thousand km (from 0.2 thousand km to 2.6 thousand km), in Siberia and the Far East – 2.2 thousand km (from 0.4 thousand km to 4.7 thousand km).

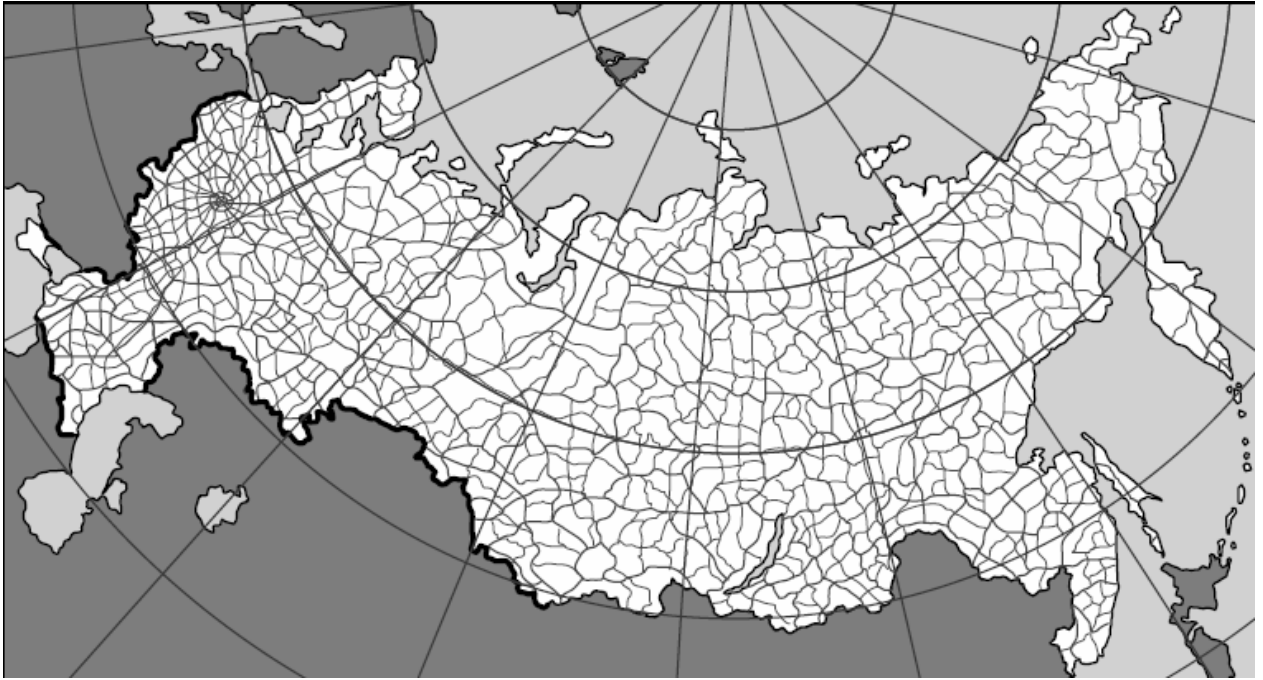

Fig. 8. Main Russian Vertical Reference Frame as of 01.01.2023

The MRVRF of the country is dense with networks of 3 and 4 classes. Levelings of classes I-IV ensure uniform distribution of the system of normal heights throughout the entire territory of Russia.

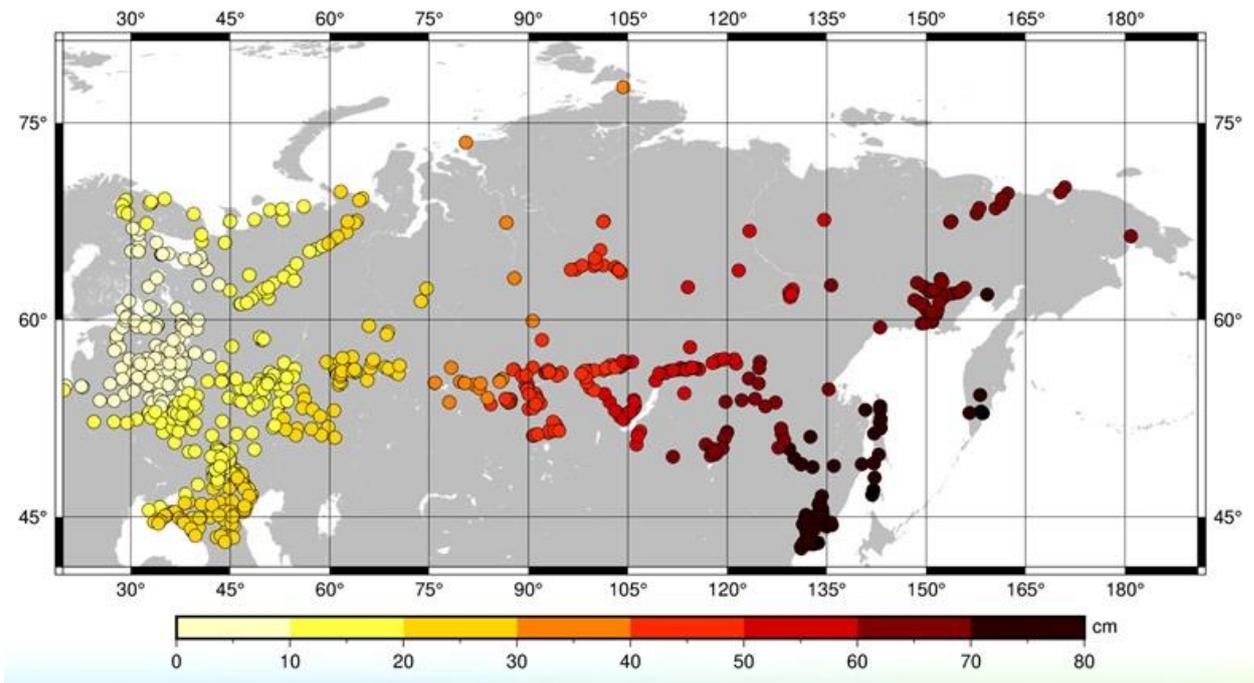

Fig. 9. Propagation of RMS errors based on the results of the general adjustment of the Main Russian Vertical Reference Frame.

The solution of the issues of further development of the system of height and gravimetric support is primarily associated with the improvement of the network of the Main Russian Vertical Reference Frame, as the basis of the state system of normal heights. This improvement, in our opinion, should go in two directions - the creation of several initial points of the Main Russian Vertical Reference Frame, with the performance of a complex of works on them from laying a group of centers (initial, working, control) to performing satellite, leveling and gravimetric measurements in connection with existing networks of MRVRF, FAGN, PGN points and points of the fundamental gravimetric network.

The work [Popadyev et al., 2021] shows the most favorable places on the territory of Russia for organizing starting points when establishing a state system of heights consistent with the world system of heights; it is indicated in the vicinity of which settlements it is convenient to choose a fundamental or secular reference benchmark, and for clarity, the radii of the near zone are given when calculating the height anomaly by numerical integration. The scheme of organization of initial points and the radii of the near zone when calculating the height anomaly is shown in Fig. 10.

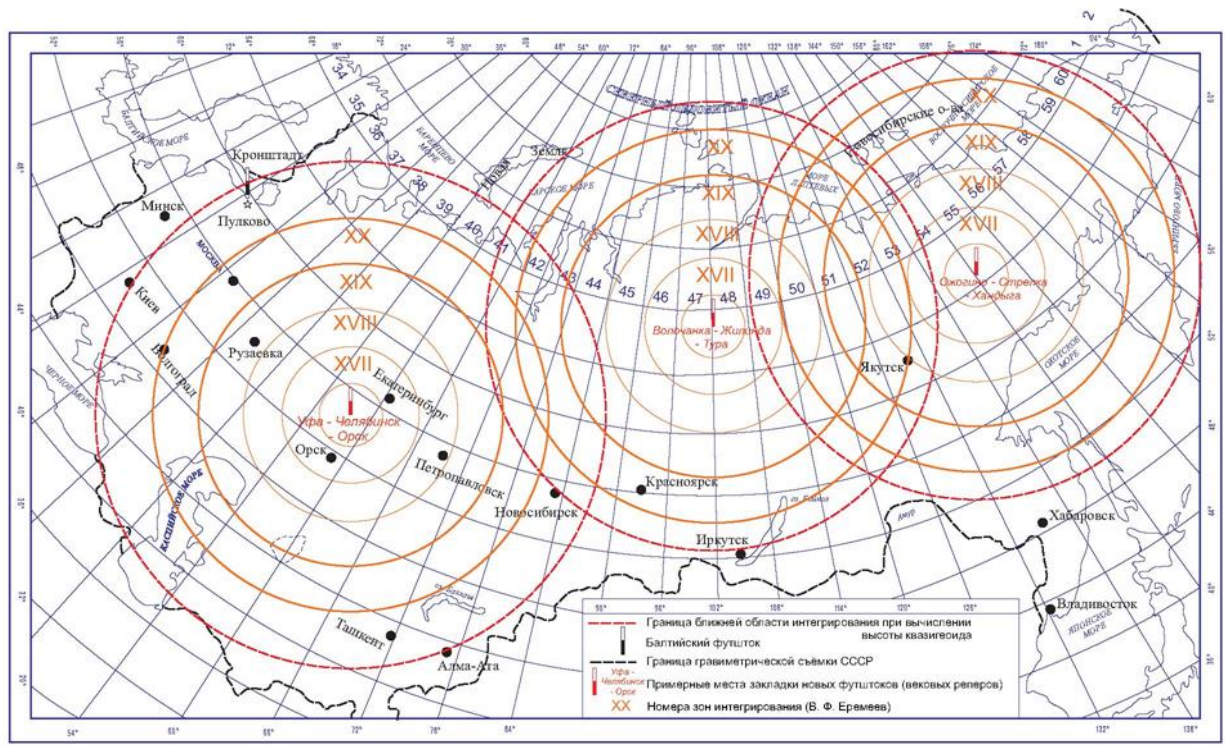

Fig. 10. Scheme of organization of initial points and radii of the near zone when calculating the height anomaly.

The State Gravimetric Network is the basis for performing gravimetric measurements aimed at studying the gravitational field and the shape of the Earth and their changes over time, as well as for solving other scientific and economic tasks, including metrological support for gravimetric surveys (see Fig. 11).

| No | Site order | Number |
|----|------------|--------|
| 1 | Initial (main) points | 2 |
| 2 | Fundamental points (complimentary sites) | 113 (117) |
| 3 | 1st order points (complimentary sites) | 579 (277) |

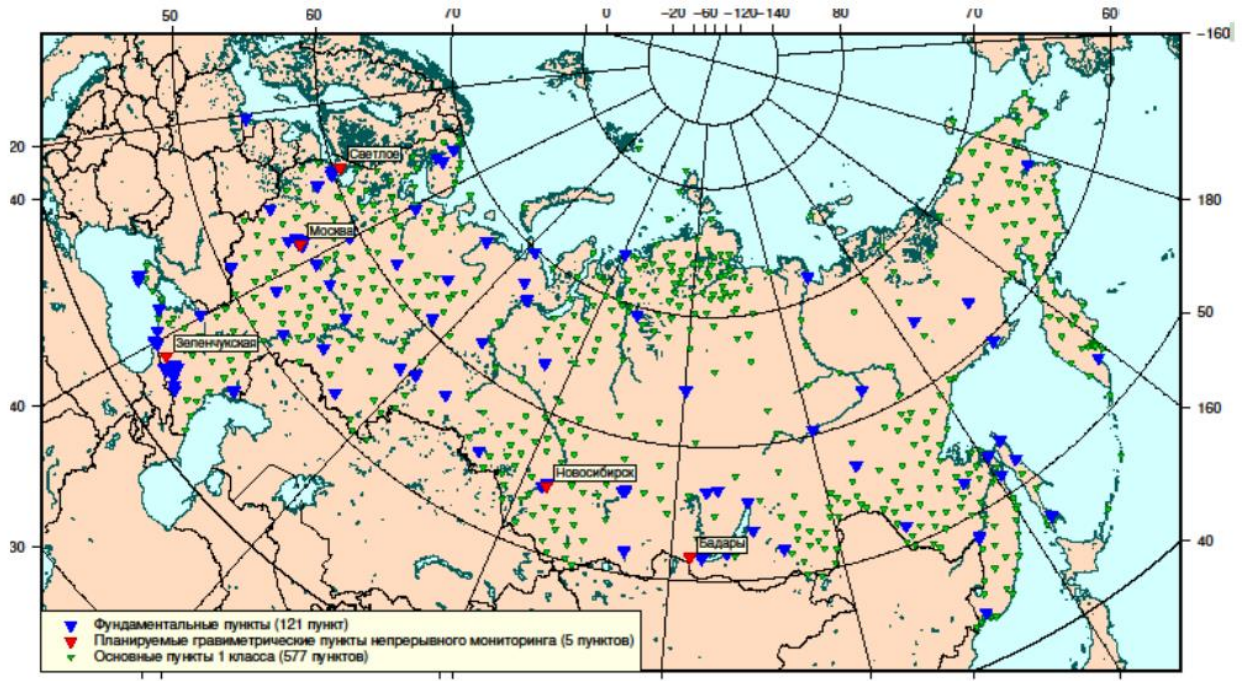

Fig. 11. State Gravimetric Network as of 01.01.2023

State geodetic works also includes observations at the geodynamic test areas of Russia located in the Caucasus, Crimea, Baikal, and Sakhalin Island. In the last century, the state controlled the movements and deformations of the earth's crust as many as dozens of geodynamic seismic and technogenic test areas. Further expansion of these studies is expected in the future.

Especial work is planned on the international project of quantum footstocks.

Research on the theory of normal heights continues, which is the subject of a number of publications [Popadyev, 2019, 2021; Popadyev, Rakhmonov, 2022a, b; Popadyev et al. 2021].

The development of the theory of adjustment of geodetic networks using the least squares method and coordinate transformation methods is considered in [Neiman, Sugaipova, 2019a, b; Neiman and Sugaipova, 2022].

In 2022, the Center for Geodesy, Cartography and SDI completed an experimental FAGN adjustment. The main material for obtaining the series of FAGN stations coordinates in the ITRF system was the results of processing and adjusting the FAGN measurements for 2015–2022. The points from the IGS core were taken as initial points, which ensure the global embedding of FAGN into the IGS framework.

The disadvantages include the fact that the adjustment is performed in a tide-free system, while on the ground it is convenient to equalize in a zero-tide system (zero-tide).

Processing and equalization were performed "in four hands".

Realizations of terrestrial coordinate systems in the 20th century have gone from catalogs of disparate astronomical and geodetic networks to permanent stations.

Depending on the level of accuracy of the tasks being solved, the results of the network adjustment can be interpreted as a coordinate system with a different, increasing level of detail:

1. Coordinates fixed "forever"; before the introduction of other coordinate systems (SK 1932, 1942 and 1995), terrestrial methods.

2. Coordinates fixed to the standard epoch. An example of implementation is the coordinates of points of the State Geodetic Network of Russia in the GSK-2011 system. Reduction to the standard epoch is carried out indirectly with the help of ITRF, in which GSK-2011 is identical to ITRF-2008 for epoch 2011.0 by definition.

3. The same + estimates of linear "velocities" of change of coordinates to bring them to some epoch close to the standard one.

4. The same + periodic nonlinear harmonic variations. Discontinuous effects can be taken into account separately in the neighborhood.

To bring the coordinates to the old epochs, linear transformation parameters defined by the developers or independently can be used.

5. Geophysical interpretations, identification of additional tectonic formations, their poles and rotation parameters, etc. These are additional ways to improve and maintain accuracy.

The obtained time series of geocentric coordinates of FAGN points and their changes in the topocentric system are smoothed using a linear-harmonic approximation, where first a linear (long-term) trend in coordinates change (for example, $X$) is selected and removed in the form:

$$X(MJD) = a \cdot MJD + b,$$

which is equivalent to the model accepted in practice

$$(MJD) = X_0 + v_X \cdot (MJD - MJD_0),$$

where $X_0$ — site position for a standard epoch, wherein

$$a = v_X, \, b = -v_X \cdot MJD_0,$$

where the long-term rate of change $v_X$ is expressed in [m/day]. As a result, the position of the point at any point in time is represented as follows:

$$(MJD) = X_0 + v_X \cdot (MJD - MJD_0) +$$
$$+ C_1 \cos \omega \, MJD + S_1 \sin \omega \, MJD +$$
$$+ C_2 \cos 2 \, \omega \, MJD + S_2 \sin 2 \, \omega \, MJD + \ldots,$$

Where circular frequency $\omega = 2\pi/365$, harmonically coefficients $C_n$, take into account intra-annual variations in the coordinates of the point. Coefficients of linear approximation of geocentric coordinates $X_0$, zero velocities $v_X = 0$, as well as harmonic coefficients of intra-annual changes in geocentric coordinates $C_n$, $S_n$ the most stable points – the source material for setting the coordinate system on the territory of the country. FAGN points in Russia are rarely subject to abrupt changes. Topocentric coordinates show pronounced periodic changes both in plan (Fig. 12) and in height (Fig. 13). On Fig. 13 harmonics from all FAGS points are superimposed, the general behavior can be traced in the high-altitude component.

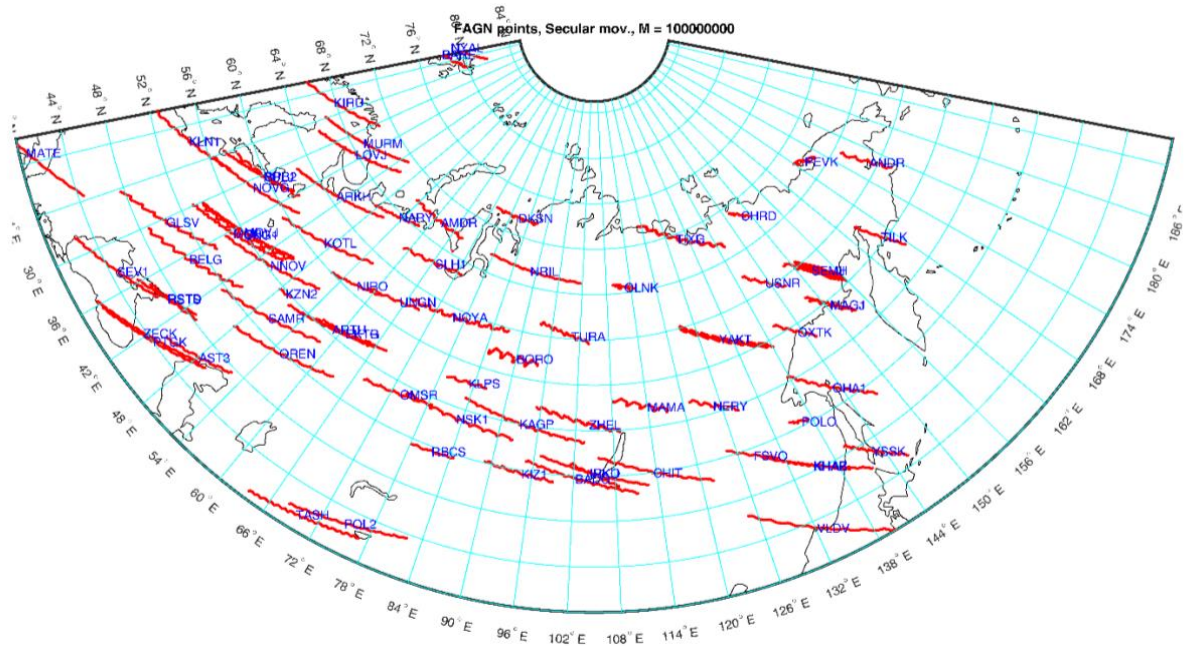

Fig. 12. Horizontal changes in the coordinates of the FAGS settlement. Red shows the trace of moving the item.

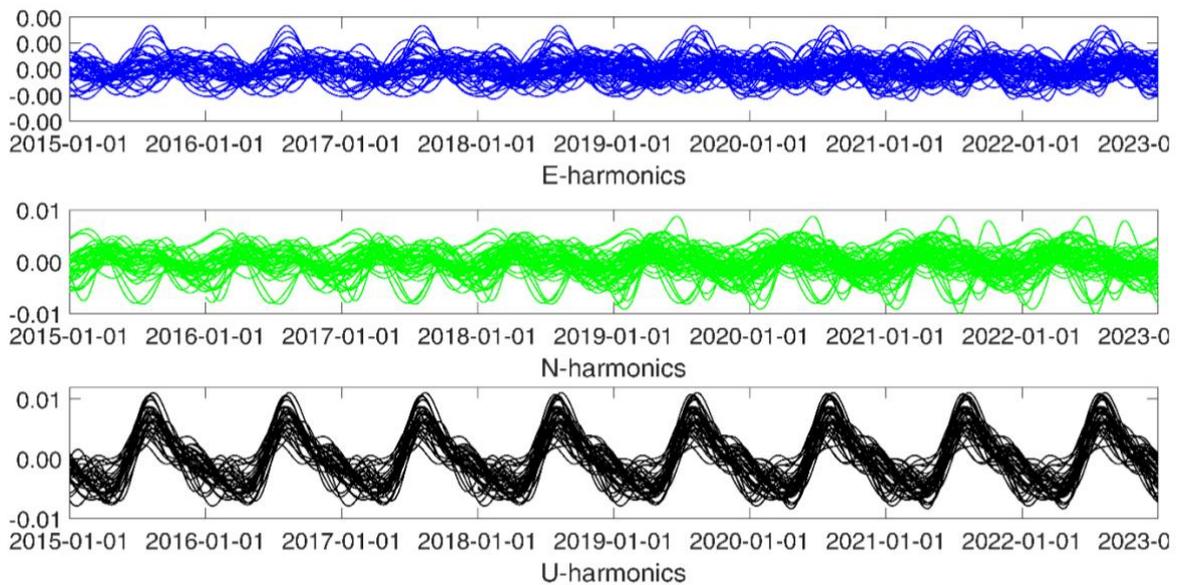

Fig. 13. Periodic changes in the coordinates of FAGN points.

# Gravity Field

Pasynok S.[1], Popadyev V.[2], Sugaipova L.[3]

[1]National Research Institute for Physical-Technical and Radio Engineering Measurements (VNIIFTRI), Mendeleevo, Moscow Reg., Russia

[2]Federal Scientific-Technical Center of Geodesy, Cartography and Spatial Data Infrastructure

[3]Moscow State University of Geodesy and Cartography, Moscow, Russia

## Gravimetry

The results of several international comparisons of absolute gravity measurements at points in the Russian Federation from 2007 to 2013 are summarized. [Yushkin et al., 2021]. An increase in gravity acceleration has been obtained instead of the expected decrease. The authors attribute the discrepancy to the influence of asthenospheric currents in this region.

The organizational issues of the application of metrology in the field of studying the gravitational field are discussed, the concepts, considers the standard of acceleration of gravity are analyzed in [Vitushkin, 2020; Vitushkin et al., 2020; 2021]. The state of the International gravimetric reference systems and reference frames are considered in the work [Wziontek et al., 2021].

## Global Studies

Preliminary results of building a geoid model within the water area of the World Ocean according to satellite altimetry data are presented in [Tsyba et al., 2020]. At present, satellite altimetry has rightfully taken its place among the methods of remote sensing of the Earth from space. The scope of satellite altimetry is constantly expanding. In addition to the problems of geodesy that have already become classical, satellite altimetry data are actively used in many geosciences. The use of satellite altimetry data to determine gravity anomalies gives good results at a relatively low cost. These advantages lead surveyors to use this method as an alternative when determining the surface of the geoid in offshore areas. The paper considers the issue of determining the characteristics of the gravitational field in the waters of the World Ocean according to satellite altimetry data. On the basis of the "delete-recovery" method, special software was developed and a digital geoid model was built for a section of the Black Sea water area (Long. 36.1°–37.76°, Lat. 42.74°–44.38°). The standard deviation from the data of the EGM2008 global model in the indicated area for gravity anomalies does not exceed 6 mGal. Based on the method of processing altimetry measurement data considered in this paper, it is planned to build a global geoid model within the water area of the World Ocean.

## Local Studies

Airborne gravimetry makes it possible to obtain gravimetric measurement information in hard-to-reach areas, complementing satellite and ground-based methods for measuring geopotential transformants. The downward continuation of airborne gravimetric measurements is one of the key tasks in processing these data. In the papers [Sugaipova and Neyman, 2022; Neyman, Sugaipova, 2022], it is proposed to use SRBF to solve this ill-posed problem. The computational experiments carried out using model data have shown that this approach allows to obtain highly accurate and stable results. In addition, the resulting approximating construction in the form of a linear combination of SRBF can be used to calculate the values of both gravity disturbance and disturbing potential, and its other transformants in the approximation region.

In [Nguyen, Neyman et al., 2022], based on the results of satellite altimetry, gravity anomalies in the waters of Vietnam and in the adjacent territory were determined with high accuracy. The accuracy of the calculated gravity anomalies in the study area is estimated by the root-mean-square error of 1.18 mGal, which better than the results of all previously performed similar works.

Evaluation of the quality of representation of gravity anomalies using the global Earth's gravity model EGM-2008 on the example of the high-mountainous territory of Tajikistan [Rakhmonov, Popadyev, 2021], the consistency of gravity anomalies from the gravimetric map of the USSR M 1 : 1,000,000 and anomalies from the global Earth's gravity model is characterized by a two-dimensional correlation coefficient of 0.94.

***Physical Geodesy***

Due to the development of global navigation satellite systems (GNSS), the problem of determining the geoid heights from the gravity disturbances is becoming increasingly actual. This problem has a well-known solution by means of Hotine-Koch integral transform. To solve the same problem using regional gravity data, Hotine-Koch kernel should be modified. In [Neyman, Sugaipova, 2020; 2021] relevant modifications of the kernel, based on the theory of Molodensky and Ostach coefficients, are proposed. The frequency characteristic of the truncation operator of Hotine-Koch kernel onto inner zone of $\psi_o$ radius is introduced. When solving this problem, it is proposed to replace the integration with an approximation by means of spherical radial basis functions (SRBFs). As SRBF, new scaling functions and wavelets are introduced that use the specified frequency response. The performed numerical experiment showed the high accuracy of the approximation results.

In the works of M. S. Molodensky in 1956 and 1960 when solving the problem of determining the anomalous potential $T$ on the basis of mixed gravity anomalies $\Delta g$, the expression for the derivative of the inverse distance $1/r$ between points on the surface of the ellipsoid $S$ in the direction of the outer normal $v$ to the surface of the ellipsoid is used:

$$\frac{\partial 1/r}{\partial v} = -\frac{1}{2Nr}\left(1 + \frac{e'^2}{r^2}(N\sin B - N_0\sin B_0)^2\right),$$

where $e'$ is the second eccentricity of the ellipsoid, $r$ is the length of the chord connecting the points on the surface of the ellipsoid with latitudes $B$ and $B_0$, $N$ and $N_0$ are the corresponding radii of curvature of the first vertical, index 0 refers to the calculation point. In this expression, given by M. S. Molodensky without derivation, the authors of [Mezhenova, Popadyev, 2022] found an error, which, however, does not affect the solutions with preservation of terms of the order of the square of the eccentricity (but affects the deductions with preservation of terms with the fourth degree of eccentricity) , the correct expression is obtained for the derivative of the function of the reciprocal length of the chord along the direction of the outer normal to the ellipsoid:

$$\frac{\partial 1/r}{\partial \nu} = -\frac{N - N_0 \cos\psi_\Gamma - e^2 \sin B (N \sin B - N_0 \sin B_0)}{r^3},$$

where $e$ is the first eccentricity of the meridian ellipse, $\psi_\Gamma$ is the angle between the normals to the ellipsoid, restored at points at the ends of the chord:

$$\cos\psi_\Gamma = \sin B \sin B_0 + \cos B \cos B_0 \cos(L - L_0).$$

Taking into account the new expression, the problem of determining the anomalous potential is solved, taking into account compression by pure gravity anomalies.

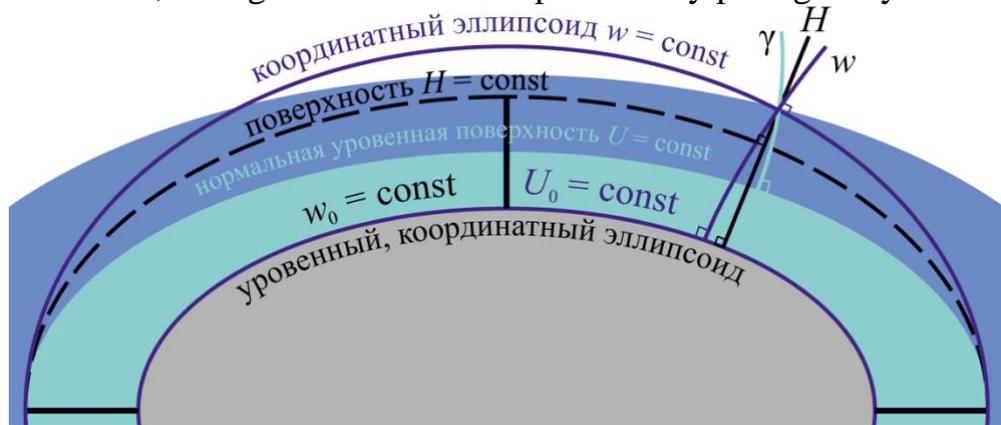

Fig. 1. Types of heights in the gravitational field of the Earth

In the field of the theory of heights in the gravitational field of the Earth, work continues to justify the application and study of the special properties of the system of normal heights arising from the theory of M. S. Molodensky. In the reviews [Popadyev 2019, 2021] on the reports of the IX Hotine-Marussi Symposium dedicated to orthometric heights, the advantages of normal heights over orthometric ones are indicated even in the unrealistic case, if the position of the geoid is exactly known (the distribution of anomalous masses is given), the main criterion is the maximum constancy of the height value on the same level surface. In the development of this topic, the most common inaccuracies and "preconceived notions" in the presentation of the theory of heights in Russian literature are disassembled [Popadyev et al., 2021], detailed comments are given on the main connections between the geoid and quasigeoid, possible illustrations with examples and the simplest interpretations of these concepts are given. Taking into account the fact that even in the specialized literature there are different indications of the method of calculating the normal height: along the normal to the reference ellipsoid (see Fig. 1), along the coordinate line of the spheroidal system, or along the force line of the normal gravity field (the most related to the orthometric system) , the advantages of calculating the normal height as the length of the coordinate line of the spheroidal coordinate system are indicated [Popadyev, Rakhmonov, 2022], working formulas are derived, inaccuracies in the expressions of M. I. Yurkina in the method of TsNIIGAiK 2004 are corrected.

# Geodynamics

Dokukin P.[1], Kaftan V.[2], Mikhailov V.[3], Shestakov N.[4,5]

[1]Peoples' Friendship University of Russia named after Patrice Lumumba, Moscow, Russia
[2]Geophysical Center of the Russian Academy of Sciences, Moscow, Russia
[3]Schmidt Institute of Physics of the Earth of the Russian Academy of Sciences Moscow, Russia
[4]Far Eastern Federal University, Vladivostok, Russia
[5]Institute for Applied Mathematics, Vladivostok, Russia

### *Crustal weather studying*

Long-term continuous GNSS observations in the geodetic networks of the world made it possible to discover interesting regularities in the relationship between the deformation process and seismicity and tectonics. The analysis algorithm consists in obtaining daily frames of accumulated deformations and seismic activity on maps of active faults. The frames are combined into a kinematic visualization, by considering which it is possible to identify patterns in the behavior of the earth's crust over a multi-year time interval [Dokukin et al., 2022a, b, c; Dokukin et al., 2023a, b, c, d]. Similar to hydro-meteorological studies, this approach forms a scientific direction that can be called geosynoptics.

The network of continuous GNSS observations CORS-TR (Continuously Operating Reference Stations – Turkey) served as the basis for studying the movements and deformations of the earth's surface in the Eastern Anatolia region. The region is an area of collision of the Nubian and Arabian tectonic plates with the Anatolian block and the Eurasian plate. It is characterized by increased seismic activity (several devastating earthquakes occurred here in 2009–2023). The main tectonic structure of the region, the North Anatolian Fault Zone (NAFZ), marks the boundary of the Eurasian and Arabian tectonic plates and, along with the Varto Fault Zone (VFZ), creates an obstacle to the movement of the Arabian Plate to the North, turning the movement of the lithosphere towards the west-north-west inside segment limited by NAFZ and EAFZ. The key object that determines the nature of tectonic movements is the Karliova triple junction area, dissected by local faults and including volcanic areas.

Based on the results of GNSS observations in the CORS-TR network, a team of Russian and Turkish geodesists and geophysicists determined the horizontal displacements of its points for each day of observations and built time series. Point displacements are presented in the local reference frame, as it is more effective for studying the mutual multidirectional movements of the sides of local faults (whereas, point displacements expressed in the global reference frame primarily demonstrate the trends in the movement of the global tectonic plate on which they are located) . According to the displacement values, digital models of the distribution of horizontal deformations were formed, for the calculation of which the horizontal deformation tensor was used. Thus, the main deformations $\varepsilon_{1,2}$, and total shear $\gamma$ and dilatation $\Delta$ were determined. In view of the significant unevenness of the finite elements, the strain values were reduced to the average area of the triangle of the network.

The obtained spatial models of movements and deformations were combined into dynamic geosynoptic maps (video animations), providing a visual analysis of the seismic-deformation process based on the fault tectonics of the region. Synoptic animations have become the basis for analyzing the spatiotemporal relationship between fault tectonics, seismicity, movements, and deformations of the study area [Dokukin et al., 2022a, b, c; Dokukin et al., 2023a, b, c, d].

Geosynoptic animation frames [Kaftan et al., 2022a] demonstrate the evolution of dilatation, anomalous changes in which may be due to discharge in the earthquake source. Heuristic analysis shows that two years after the first strong earthquake M6.1 (2010), significant compressive deformations began to develop in the region of the Varto fault zone, and a year later, an extremum of spatial extension formed in the epicenter region, which began to develop without noticeable seismic activity. At the same time, the extension area expanded to the main seismic event M6.7, 2020, which occurred at the boundary of the interaction of compression and extension regions, revealing the shear mechanism of seismic rupture. Later, a series of moderate Bingol earthquakes also occurred in the north of the Karliova Triple Junction, on the boundary of interaction between the main extrema of extension and compression. The revealed evolutions of deformations make it possible to consider the Karliova Triple Junction as a source of stress propagation that determines the kinematics of the East Anatolian block.

In regions with a predominance of the shear mechanism of movement along active faults, slow deformation shear waves are found [Kaftan, Melnikov, 2019; Kaftan and Tatarinov, 2022b]. A heuristic analysis of geosynoptic animation of full shear deformation in the Eastern Anatolia region [Kaftan et al., 2022b] shows that a year before the M6.1 (2010) earthquake, a significant deformation extremum formed in the region of the future epicenter, which allows us to conclude that a strong seismic event is caused by the accumulation of shear deformations that contribute to the release of stresses within the mature seismic source. Visual analysis makes it possible to observe the further seismic-deformation process and draw conclusions about the mechanisms of stress accumulation and release during strong earthquakes, as well as the role of the Karliova Triple Junction as the intersection of the main fault zones of the region. Attention is drawn to the sequence of events of the Elazig mainshock on January 24, 2020 M6.7, as well as to the peculiarity of deformation propagation from the first strong mainshock parallel to the NAFZ and EAFZ shear directions.

Having long-term data on the nature of the movements of the earth's crust in the study area, the authors analyzed the trends in the accumulation of the module of motion vectors of GNSS observation points. The corresponding geosynoptic animation is presented in [Dokukin et al., 2023]. The first strong seismic event M6.1 (2010) occurred in the zone of internal displacement deficiency expressed in dark brown color, then there was an increase in internal displacements, which accompanies the release and relaxation of elastic stresses. The deficit area shifted in two directions to the southwest and northwest along the distribution of NAFZ and EAFZ, where four moderate earthquakes with M~5 later occurred. Two of them, located southeast of the EAFZ, caused an increase in the mobility of the earth's crust at the epicenter of the strongest earthquake in 2020 (M7.1), occurred in the zone of accumulated displacements at a level of 5-6 cm and was accompanied by a co-seismic displacement of up to 13 cm A series of destructive Bingol earthquakes (2020) also occurred in the mobile zone. In the subsequent period, the entire study area was subjected to accumulated displacements of at least 5 cm, which indicates the release of elastic pre-seismic stresses in the entire area.

About a year after the completion of the research, a series of catastrophic Karamanmarash earthquakes occurred in the East Anatolian region (February 2023, M=7.5–7.8). The synoptic animation [Dokukin et al., 2023d] showed the accumulation of displacement deficit exactly in the area of these earthquakes (Fig. 1).

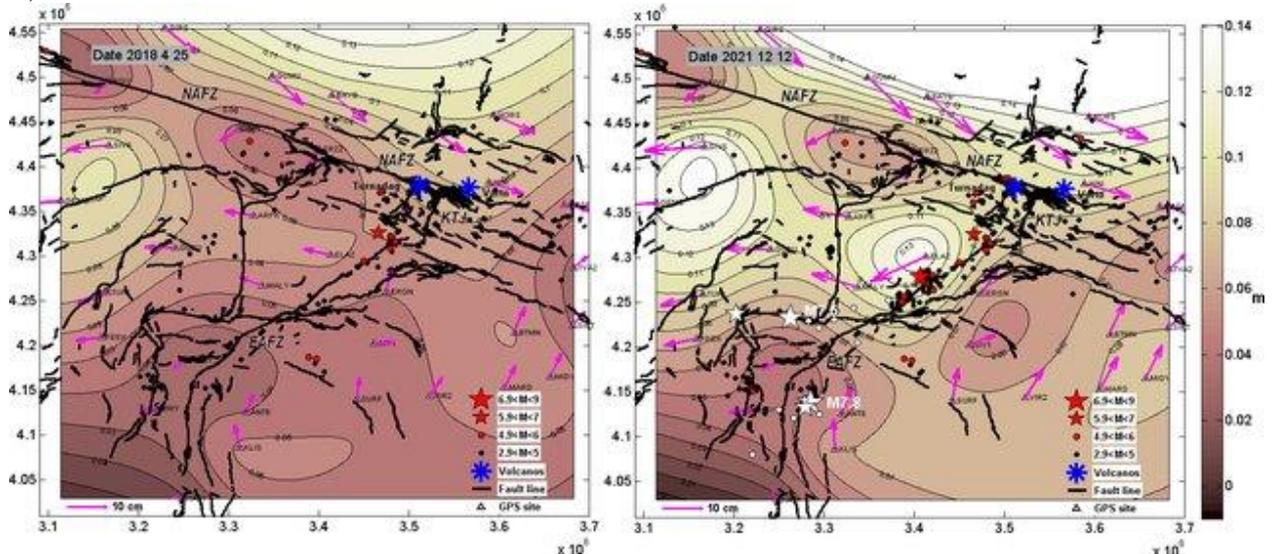

Fig. 1. Displacement deficit before the Karamanmarash 2023 earthquake series (white stars). On the left is the accumulation of displacement deficit 5 years before the events. On the right is the displacement deficit about a year before the main events (dark area in the southwest of the region).

A retrospective analysis of the evolution of the seismic-deformation process before strong earthquakes made it possible to detect the phenomenon of slow migration of total shear deformation [Kaftan, 2020, 2021; Kaftan, Melnikov, 2019; Kaftan, Tatarinov, 2022a, b]. Total shear deformation waves were detected before the strong Napa (2014, $M_w$ = 6, USA, California) and Ridgecrest (2019, $M_w$ = 7.1, USA, California) earthquakes. Anomalous total shear deformations were discovered about 7 years before the seismic events. Their propagation demonstrated the character of a soliton wave. In the first case, the anomalous deformation was initiated by a pair of moderate M5 earthquakes. It spread in a linear flow at a speed of 20 km/year. In the second case, the growth of the total shear deformation had a circular concentric character. The propagation velocity of an almost circular front was 7 km/year. The value of the deformation of the wave front upon reaching the epicentral zones and releasing the accumulated elastic stresses was $0.2-0.3*10^{-5}$. The spread of significant deformation anomalies had a triggering effect on the sources of strong earthquakes.

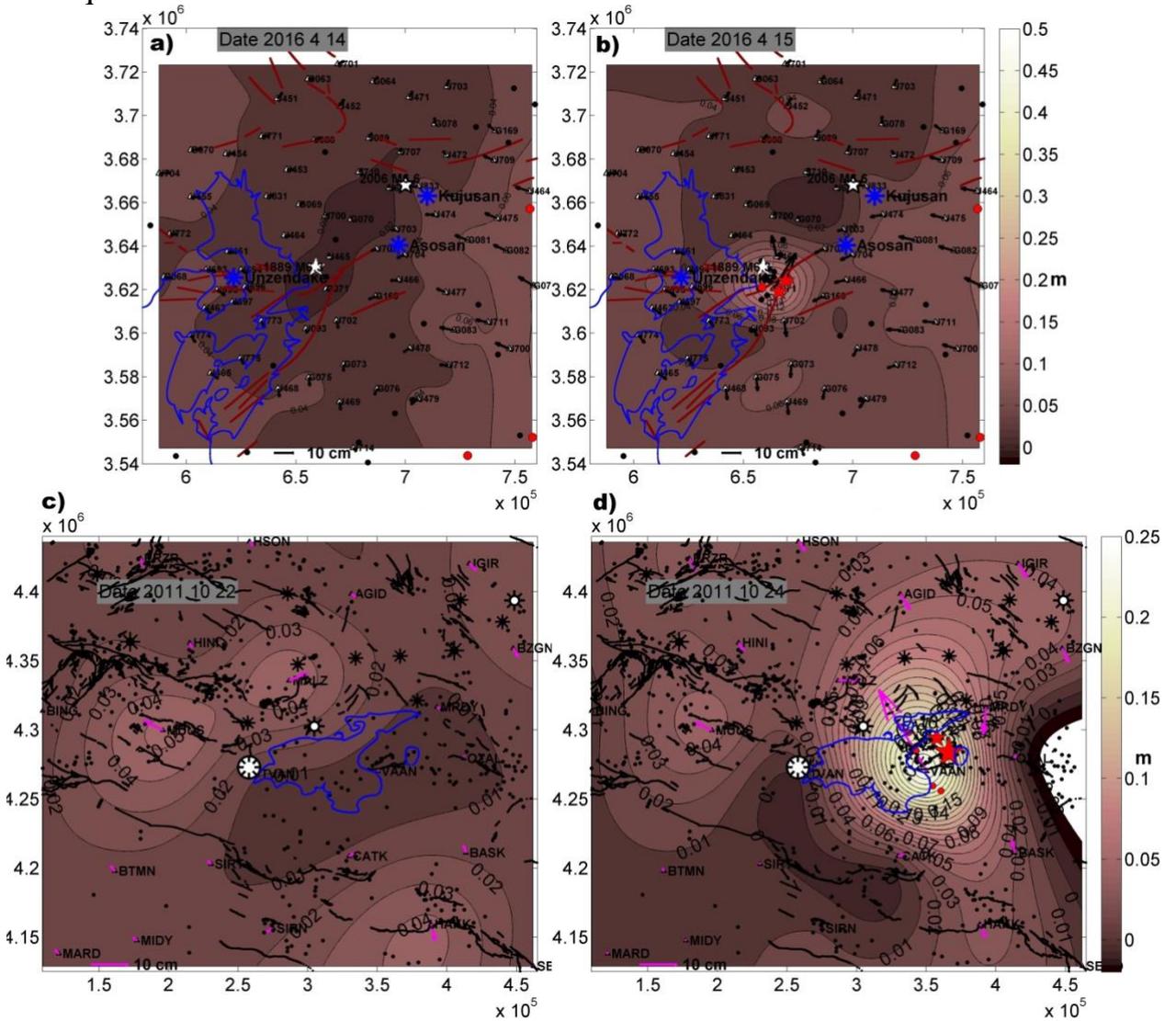

Fig. 2. Distribution of internal displacement deficit before (a) and after (b) the Kumamoto earthquake (Japan), before (c) and after (d) the Van earthquake (Turkey). The maximum deficit is dark brown areas. The maximum shifts are light areas. White stars are historical strong earthquakes. Red stars are the studied strong earthquakes. Black dots are weak earthquakes. Black lines are active faults. Even and purple arrows are horizontal displacement vectors.

In the seismically active Parkfield area (USA, California), a quasi-linear full shear wave was detected in the San Andreas fault zone in the absence of strong seismic events. In this zone, strong earthquakes M ≥ 6 occur with a frequency of about 20-30 years. The wave propagation velocity was 2 km/year. Since the periodicity of strong earthquakes indicates a high probability of the occurrence of the next event, the result obtained is of great importance for estimating the waiting time, as well as the location of a future strong earthquake.

In recent years, in studies of the evolution of movements and deformations of the earth's surface, attention has been drawn to the search for closed zones of seismically active faults and the assessment of the deficit of "slip" along these faults. Such zones are considered as potential sites of strong earthquakes. Geodetic networks do not always register movements along faults due to their insufficient density. In this case, to identify "locked zones", it is convenient to characterize the deficit of accumulated internal displacements (movements) [Gvishiani et al., 2020a, b]. Examples are shown in Fig. 2.

Areas of motion deficiency, in areas of future strong earthquakes, were retrospectively obtained from GNSS observations. Before the Van earthquakes (M7.2, 2011, Turkey) [Kaftan et al., 2021a, b] and Kumamoto (M7.3, 2016, Japan) [Kaftan et al., 2022a, b] over approximately 3 and 7 years, respectively, zones of displacement deficiency were accumulated. The epicenters of the above Napa (2014, M6, USA, California) and Ridgecrest (2019, M7.1, USA, California) earthquakes also fell into extended areas of minimal movements. Empirical evidence is accumulating that it is possible to predict the locations of future strong earthquakes from geodetic observations.

### *Regional geodynamics study*

Particular attention to the control of tectonic movements and deformations is drawn to the area of the future disposal of radioactive waste [Gvishiani et al., 2020a, b; Gvishiani et al., 2021a, b, c, d; Gvishiani et al., 2022a, b; Manevich et al., 2021a, b, c; Tatarinov et al., 2022].

The problems of modern geodynamics of the northeast of Asia, which includes the northeast regions of China and the Far East of Russia, are considered in the papers [Meng et al., 2019; Yu et al., 2019; Bykov et al., 2020]. A brief review of the development stages of the Unified Geodynamic Observation Network of the Far East Branch of the Russian Academy of Sciences, the main results of seismological and GPS/GLONASS observations obtained within the framework of the target complex scientific research program of the Far East Branch of the Russian Academy of Sciences "Modern geodynamics, active geostructures and natural hazards of the Russian Far East (2009–2013))" and projects of the FEB RAS in 2014, 2018, 2019, as well as the achieved positions of the FEB RAS in the field of geodynamics is given. According to the data of continuously operating and periodically observed stations of Chinese and Russian GNSS networks, for the first time the parameters of modern movements and deformations of the earth's crust of the study region after the Great Tohoku earthquake on March 11, 2011, Mw 9.1, were obtained, including coseismic displacements and postseismic changes in the secular velocities of movements of geodetic points. Postseismic changes in the stress-strain state of the lithosphere and seismic activity near the largest fault structures in the region: the Ilyan-Itun and Dunhua-Mishan faults are analyzed.

### GNSS tsunami warning

Currently, GNSS systems are widely used the paper of earthquake and tsunami early warning systems. The possibilities of using GNSS to build such systems in the Kuril-Kamchatka region, as well as problems and methods for improving the accuracy and efficiency of inversion of coseismic displacements of the seabed and land, initiated by an underwater tsunamigenic earthquake, are considered in the papers [Nechaev et al., 2020a, b; Pupatenko et al., 2021]. On the basis of the numerical experiments carried out, estimates of the accuracy and reliability of determining the parameters of an earthquake depending on the magnitude, number and relative position of GNSS stations were obtained. It is shown that the determination of all the main parameters of the earthquake source model with a magnitude of 7.4 and higher using GNSS data is quite possible, although with not very high reliability [Pupatenko et al., 2021]. On the basis of numerical experiments with the sources of real and synthesized tsunamigenic earthquakes, it was determined that the inversion of data on coseismic displacements of the GNSS network, even with a simple one-plane source model, in many cases, is able to satisfactorily describe the real source, and the initial movements of the seabed corresponding to the inverted sources quite accurately repeat the real displacements. The accuracy of determining displacements of 3 cm horizontally and 5 cm vertically is sufficient for an adequate assessment of the earthquake parameters. Thus, it is confirmed that the use of GNSS methods and the use of the existing regional GNSS network can significantly improve the reliability of tsunami warning systems with the current capabilities of GNSS technologies [Nechaev et al., 2020a]. The accuracy of modeling the Earth's crustal deformations caused by earthquakes depends not only on the detail of the seismic event source model, but also on the complexity and structure of the Earth model used. Accordingly, when solving the inverse problem (determining the parameters of an earthquake as a result of inversion of coseismic displacements), the model of the internal structure of the Earth used also affects the final result. In [Nechaev et al., 2020b] the influence of the Earth model adopted for calculations on the values of coseismic displacements in their direct modeling is considered. The effects caused by the influence of layering, sphericity, and both factors simultaneously are considered. Three hypothetical sources with magnitudes $M_w$ 7.5 were used in the simulation; 8.0 and 8.5 belonging to the Kamchatka subduction zone. Comparison of the displacements of the seabed and points of geodetic networks obtained by numerical methods suggests that taking into account the Earth's layering and sphericity leads to a decrease in the values of the calculated displacements. It is concluded that neglecting the layering and sphericity of the Earth leads to an underestimation of tsunami heights.

### GRACE-GNSS integration

Works were carried out to study the various stages of the seismic cycle according to satellite gravimetry and geodesy data. Based on the joint inversion of GRACE satellite data, offsets at GPS sites and bottom transponders, as well as displacement fields from satellite radar interferometry data, a new model of the rupture surface of the Tohoku earthquake, Japan, March 11, 2011 was constructed [Mikhailov et al., 2019a, b]. Using data on temporal variations in the GRACE gravity field coseismic and postseismic processes in the region of large earthquakes were studied: in the Wharton basin in 2012 [Diament et al., 2020], the Maule earthquake in Chile on February 27, 2010 [Mikhailov et al., 2020]. Models of the rupture surface of the Khuvsgul earthquake in Mongolia on January 12, 2021 [Timoshkina et al., 2022], the Near Aleutian earthquake on the Commander Islands on July 17, 2027 [Mikhailov et al., 2022], and the Near Ozernovsky earthquakes in Kamchatka [Mikhailov et al., 2022] were constructed based on SAR interferometry and GPS data. For all events, models of the fault surface and displacement fields on them were built, relationships with fault tectonics and source zones of previous events were studied.

Possible causes of postseismic displacements are investigated using data from the earthquake in Chile on February 27, 2010. To estimate the contribution of the viscoelastic relaxation of stresses that arose in the lithosphere and upper mantle as a result of an earthquake to the observed postseismic displacements, numerical simulation of this process was performed using the constructed seismic rupture model. The rates of displacement of the earth's surface depend mainly on the accepted value of the viscosity of the asthenosphere. The comparison of the calculated and measured displacements at low viscosity of the asthenosphere shows that if the displacement measurements are made at a large distance from the fault surface, as is the case for earthquakes in the ocean-ocean subduction zone, then the observed displacements can be explained by the process of viscoelastic relaxation with low viscosity of the asthenosphere. In those cases where there are data on displacements above the rupture surface, as is the case for the Maule-2010 earthquake, it is not possible to explain the observed displacements by the stress relaxation process at any values of viscosity. Thus, the hypothesis of low viscosity of the asthenosphere in the area of the Maule-2010 earthquake, as in other areas of the subduction zones, is not confirmed. Therefore, the observed displacements should be associated with the processes of postseismic creep, both on the surface of the seismic fault itself and on its continuation along the strike and in depth. Similar results were obtained for the Tohoku-2011 and Sumatran-2004 earthquakes.

### SAR interferometry

A number of works are devoted to the search for and evaluation of displacement fields in the areas of active volcanoes on the Kamchatka Peninsula using SAR interferometry methods. Various numerical models have been developed to interpret the obtained strain fields. For example, when studying the Tolbachik lava field, which was formed as a result of the 2012–2013 eruption, an area with an anomalously high rate of surface subsidence was identified using SAR data [Mikhailov et al., 2020; Volkova et al., 2022]. A new mathematical model of lava flow cooling was constructed, which made it possible to obtain estimates of various physical properties of lava (the content of pores, glasses), the rate of formation of the flow thickness, and also to explain the reasons for the appearance of an area with an abnormally rapid subsidence of the lava surface. Similarly, when studying pyroclastic deposits formed as a result of the eruption of the Shiveluch volcano on August 29, 2019, the rate of subsidence of a layer of pyroclastic deposits was determined using SAR interferometry using a series of Sentinel-1A images. [Volkova, Mikhailov, 2022]. To study the processes affecting the shrinkage of the material erupted from the interior of the volcano, a thermomechanical model was constructed, which takes into account the compaction of deposits due to changes in their porosity and density over time, which made it possible to estimate the physical parameters of the pyroclastic at Shiveluch volcano and explain the mechanism of vertical deformation during its cooling down.

With the help of SAR interferometry, not only surface processes are studied, but also internal processes associated with volcanic activity, with the movement of magma. During the Koryaksky volcano eruption in 2008-2009, it was extremely important to determine whether magma uplifted to the surface of the volcano or not, since this is a different degree of threat to the population. Additional research using SAR interferometry has contributed to answering this question. ALOS-1 satellite images revealed displacements on the northwest slope of the volcano, which were interpreted using the Okada model as a dike intrusion into the volcanic edifice [Mikhailov et al., 2021]. The volume of the intruded dike is consistent with the estimates of works describing the modeling of magma injection into a fracture, as well as works in which calculations of the eruption energy were performed.

The Bolshaya Udina volcano can also serve as an example, when some researchers associated the observed seismic activity in 2017 with the possible uplift of magma to the reservoir under the Uda structure and, as a result, assumed a high probability of an imminent eruption. However, further analysis of the displacements of the area of aftershock activity in time and the displacement fields on the slopes of the volcano, obtained in the process of processing of SAR images, showed that seismic activation most likely accompanied the process of retreat and subsidence of the magmatic melt downward from the volcano [Senyukov et al., 2020].

SAR interferometry is also an indispensable tool for monitoring areas associated with the development of mineral deposits. In [Babayants et al., 2021], a study was made of soil subsidence caused by mining operations in the area of Berezniki city. The situation in this area changed dramatically in 2006, when the world's largest accident in the development of water-soluble ores occurred – the mine of the First Bereznikovsky Potash Mine Administration was flooded. The mines are located directly under the city of Berezniki, the second largest city in the Perm Kray with a population of about 150 thousand people. The subsidence of the earth's surface, which is currently taking place in the city of Berezniki and in the adjacent territories, is steadily recorded by differential SAR interferometry using pairs of TerraSAR-X satellite images taken with an interval of 11 days, which makes it possible to monitor displacements with a minimum time delay. The calculation results allow us to conclude that, despite the continued subsidence of the earth's surface over the sinkholes in the study area, the measures taken to stabilize the process are generally effective. The obtained estimates of subvertical displacements are close to ground geodesy data, but the use of remote satellite methods makes it possible to monitor the development of subsidence over vast areas and estimate displacements in areas where ground geodetic measurements are not carried out.

Various approaches have been developed for the use of SAR interferometry for localization and monitoring of areas of active surface deformations (landslides, subsidence of soils, etc.) in the territory of Greater Sochi [an example, Mikhailov et al., 2014; Smolyaninova et al., 2018, 2019, 2020, 2021, 2022]. SAR images from satellites ALOS-1 (track 588A, 2007-2010), Envisat (orbit 35D, 2011-2012) and Sentinel 1A (orbit 43A and 123D in the period 2015-2021, as well as orbits 145A and 21D in 2018-2019 years) were used. During SAR image processing, various technologies were used: the DInSAR method implemented in the free SNAP package, various image time-series processing technologies: the PS-InSAR persistent scatter reflector method implemented in the free StaMPS/MTI package, and the PS-InSAR and small baseline SBAS methods implemented in the ENVI SARscape package. v.5.3. Based on all sets of images, maps of surface deformation rates were constructed for various areas of Greater Sochi for the corresponding time periods

SAR interferometry methods make it possible to identify with a high degree of detail areas of active surface deformations: landslides and soil subsidence in the coastal areas of Greater Sochi, where, on the one hand, a high building density makes it difficult to fix slow surface deformations by field methods, but, on the other hand, creates extremely favorable conditions for reflecting radar signal. Maps of active deformations of the surface, built as a result of interferometric processing of images, make it possible to determine the geometry of landslides, their activity, and identify new landslide areas or places of activation. These maps can be successfully used to refine existing maps of landslide manifestations in the Greater Sochi area. [Smolyaninova et al., 2019, 2022]. Time series of displacements of individual resistant scatters make it possible to evaluate the dynamics of displacements of landslide bodies over time, to determine the relationship between the activation of landslide structures and subsidence with various factors: precipitation, construction work, etc. [Smolyaninova et al., 2021, 2022].

12, 2021 From InSAR Data, Izvestiya, Physics of the Solid Earth, Vol. 58, No. 1, 2022, pp. 74–79.

# Earth's Rotation

Z. Malkin[1], S. Pasynok[2], L. Zotov[3]

[1]Pulkovo Observatory, Saint Petersburg, Russia

[2]National Research Institute for Physical-Technical and Radio Engineering Measurements (VNIIFTRI), Mendeleevo, Moscow Reg., Russia

[3]Sternberg Astronomical Institute, Moscow University, Moscow, Russia

The current state of work of the Main Metrological Center of the State Time and Frequency Service is considered in the work [Pasynok et al., 2019, 2021, 2022a, b]. MMC SSTF (NIO-7 VSUE «VNIIFTRI») in the Earth's orientation parameters (EOP) evaluation field executes the following functions: function of Main metrological center of State Service for time, frequencies and EOP evaluation (it states in Government Council of Russian Federation № 225 with changes and additions); function as Analysis Center (AS) of SSTF; function as Data Center (DC) of the satellite measurable network of metrological control sites of ROSSTANDART; function as measurable sites for EOP evaluation purposes. Besides these, NIO-7 takes part in a development works of FTP «GLONASS» for improvement of system of EOP evaluation and prediction. The results of MMC SSTF activities in the last years for EOP evaluation are considered.

The operative Earth's orientation parameters (EOP) evaluation activity has been carried out in Main Metrological Center of State Service for Time, Frequency and EOP evaluation (MMC SSTF) from time of the FSUE VNIIFTRI creating (from 1954). The role NIO-7 of VNIIFTRI as MMC SSTF is stated by government decree №225. Besides this, VNIIFTRI takes participation in SSTF activity as a source of measurable data and as an Analysis Center (AC).

In 2022 combined EOP values were formed by combination of nine independent EOP series which were obtained in IAA RAS, JSC TSNIIMASH and MMC SSTF. Using IAA RAS VGOS antennas data allows significantly increased accuracy of combined UT1 values and accuracy of UT1 prediction. The formation of operative EOP information by combination and dissemination of EOP information are provided in daily mode. Besides this, VNIIFTRI give the contribution in national and international data base by the transfer of GNSS and SLR data of observations which are provided in VNIIFTRI and its branches which are situated in Novosibirsk, Irkutsk, Khabarovsk and Petropavlovsk-Kamchatskii. The processing of every geodetic technique (GNSS, SLR and VLBI) observations data is provided in MMC SSTF in daily mode for EOP evaluation. The combined EOP data were used for analysis of the Earth's rotation angular velocity, see Fig. 1.

A number of pilot activities are under way to improve methods and tools for processing and analyzing measurement data of various types is carried out in MMC SSTF. The evaluation of orbits and corrections of spacecraft clocks, as well as the processing of satellite altimetric measurements are also carried out.

The work on the EOP evaluation is carried out at a high scientific and technical level. In order to improve accuracy and efficiency of EOP evaluation, work is planned both to improve methods and tools for processing and analyzing measurement data, and to improve the equipment of measuring sites.

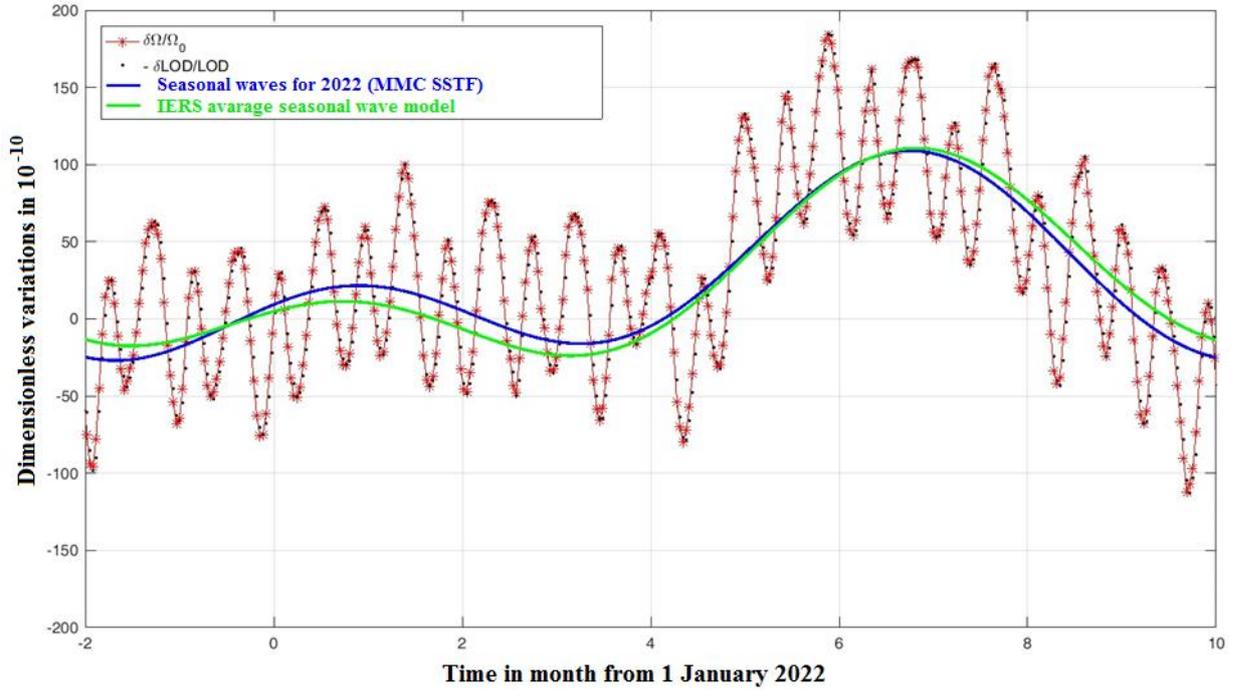

Fig. 1. Dimensionless variations of Earth's angular velocity and length of day.

Currently, FSUE "VNIIFTRI" and its East Siberian branch have installed laser systems of a new generation capable of providing millimeter accuracy in slant range measurements. In particular, in addition to the tasks of determining the PVD, it provides the ability to carry out laser transmission of time and measure the divergence of onboard time scales. However, the conditions of observation may limit the accuracy of the results obtained. It is possible to reduce, and in some cases, completely eliminate the influence of negative factors using the proposed algorithms. The introduction of the presented methods, based on the metrological capabilities of FSUE "VNIIFTRI" and proven methods for calibrating and verifying satellite laser ranging stations, allow us to rise to a qualitatively new level [Bezmenov et al., 2021].

An increase in the accuracy of determining the parameters of the Earth's rotation by combining various types of measurements is reported in the work [Pasynok, 2022]. The results of improvement of programs and methods of Earth's orientation parameters (EOP) combination of vary long baseline interferometry (VLBI), global navigation satellite systems (GNSS) and satellite laser ranging (SLR) in Main Metrological Center of State Service for Time, Frequency and Earth's orientation parameters evaluation are considered.

Nowadays, the combination of measurements made with different geodetic techniques in Main Metrological Center of State Service for time, frequency and Earth's orientation parameters evaluation is provided both at the time series level, and on the observations level. The increasing accuracy of Universal time UT1 was caused by using in combination observational data from new thirteen-meter antennas of the two-element radio interferometer, which was created in Institute of Applied Astronomy of Russian Academy of Science, see Fig. 2.

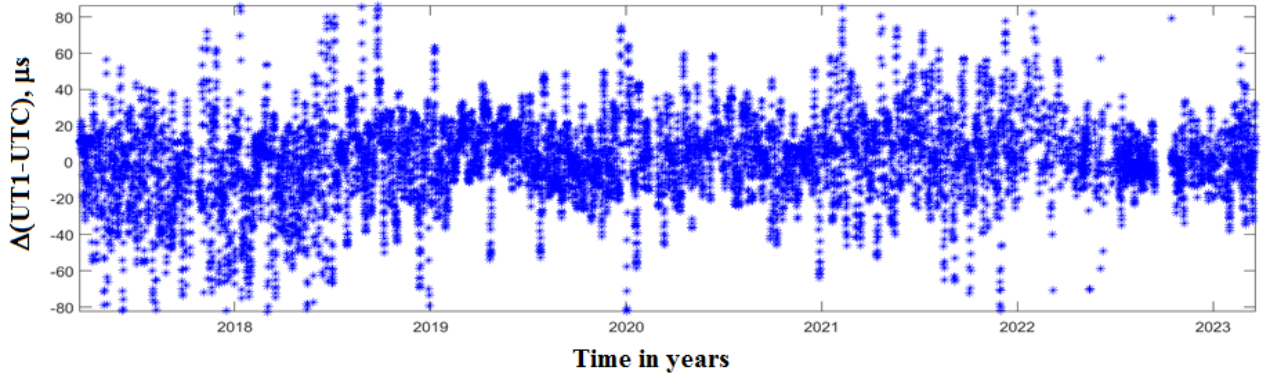

Fig. 2. The discrepancies between UT1–UTC values of EOPC04 series and results of post correlation processing in MMC SSTF of the IAA RAS correlation data [Surkis et al., 2017] for IAA VLBI VGOS network.

In general, increase in accuracy of combination values of the Earth's orientation parameters was caused both by the increase in accuracy of the separate time series, and by the improving of the combination algorithms.

The highly accurate position of the celestial pole is necessary for many applications, such as space navigation, global navigation satellite systems, and operational time determination. Most of these tasks require knowledge of the coordinates of the celestial pole, defined by the angles of precession-nutation, in real time, or even with prediction. On the other hand, the most accurate values of the precession-nutation angles are derived from VLBI observations on global networks of stations, the results of which are available with a delay of one to several weeks. Therefore, all real-time applications use the predicted positions of the celestial pole. In paper Malkin [2020a], the prediction accuracy of the celestial pole coordinates is assessed on the material on the real predictions made from 2007 at the US Naval Observatory, which functions as the Rapid Service and Prediction Center of the International Earth Rotation and Reference Systems Service (IERS), and at the Pulkovo Observatory with the prediction length up to 90 days. The two sets of predictions were analyzed and compared for both whole 2007–2016 interval of dates and for five 2-year sequential intervals. Conclusion of previous similar studies carried out in 2009 is confirmed on significantly higher accuracy of predictions made in the Pulkovo Observatory, which is about two times better for both short and maximum prediction length. It was found that the prediction accuracy did not show a clear improvement within the last 10 years, see Fig. 3.

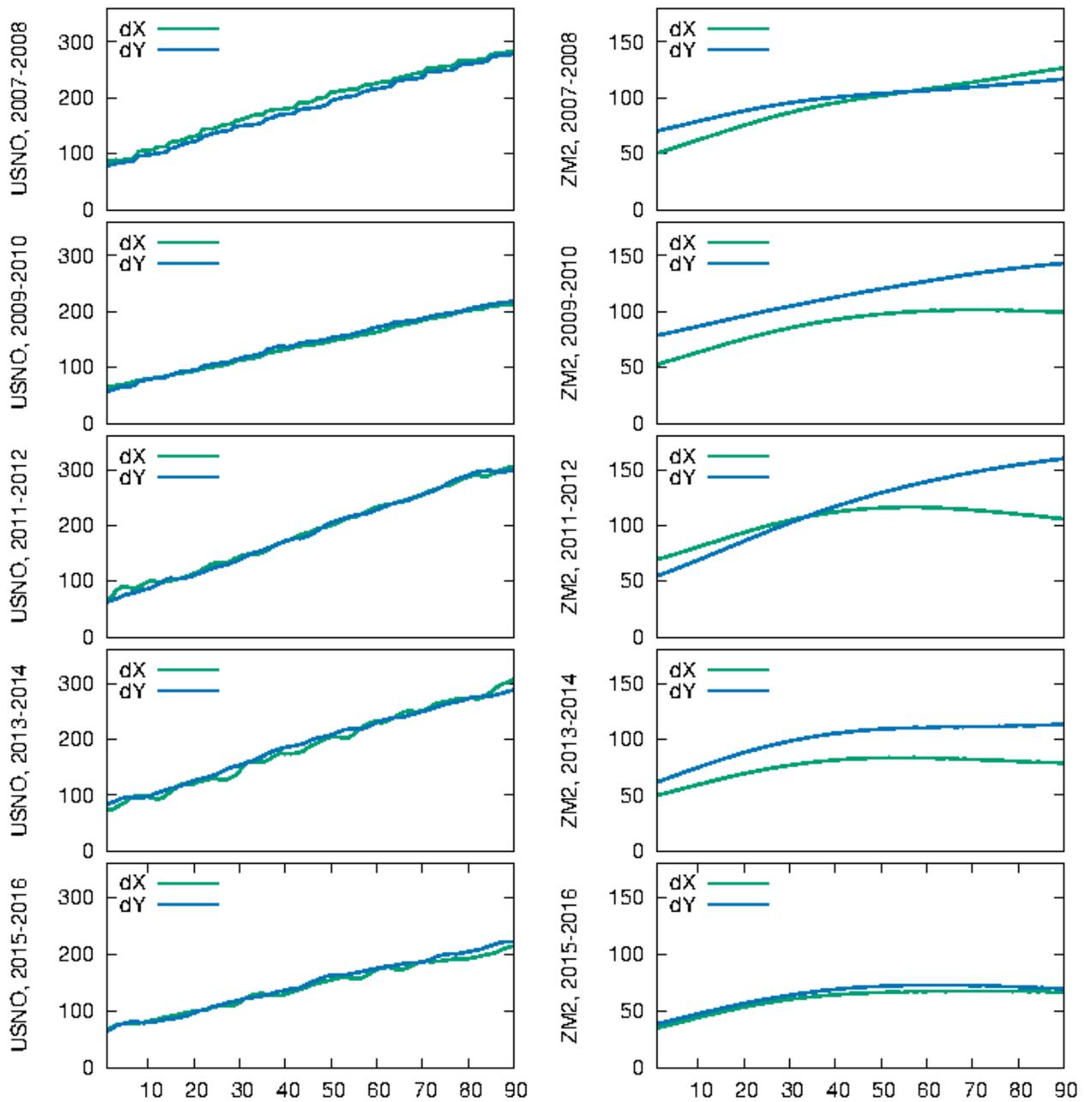

Fig. 3. The accuracy of predictions of the Celestial pole offset computed at USNO and Pulkovo Observatory (ZM2).

It was shown in several earlier studies that the results of optical astronomical observations of the time and latitude variations can be affected by earthquakes. Such an impact, if confirmed, can be used for earthquake prediction. In Hu et al. [2020] the connection was investigated between the non-polar time and latitude variations computed as the residuals between the observed values and those computed from the Earth rotation parameters provided by the IERS, and earthquakes occurred in vicinity of the Yunnan observatory in 2010–2016. A possible correlation of the jumps in the observed time and latitude time series with large earthquakes in the vicinity of the observatory is presented in Fig. 4.

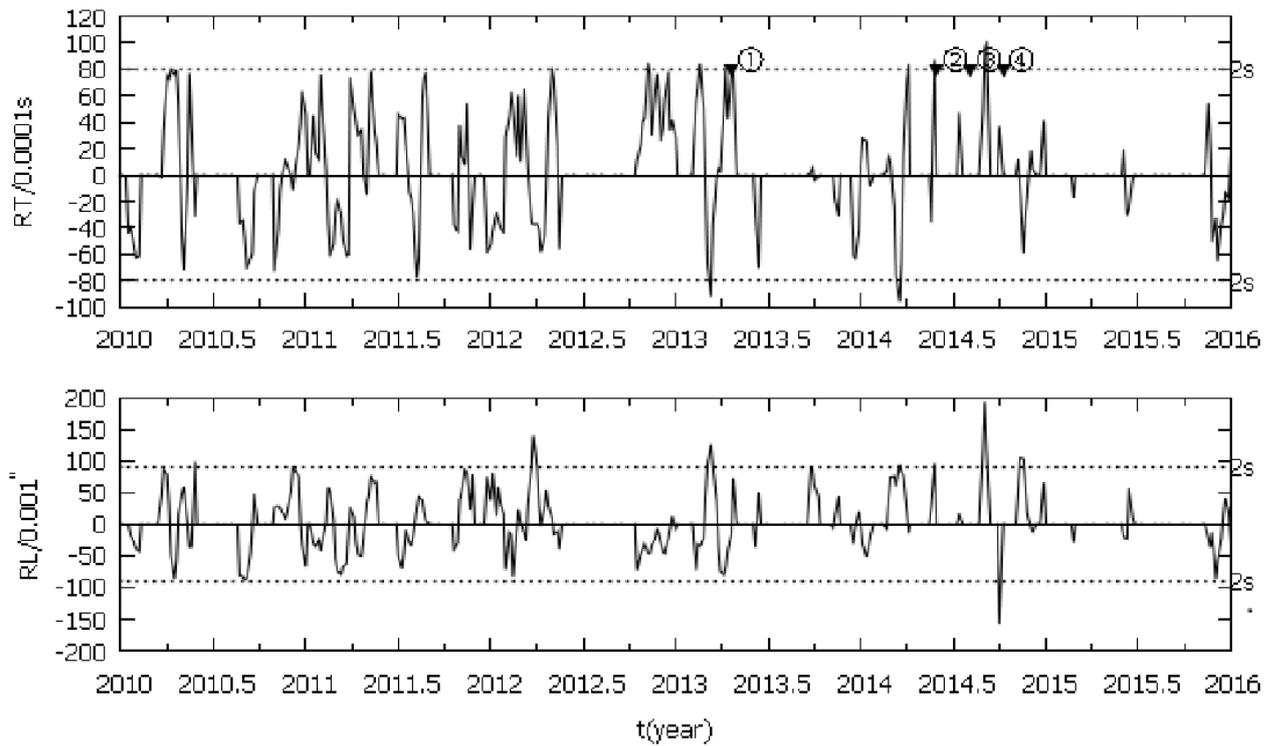

Fig. 4. Residuals of astronomical time (top) and latitude (bottom) observations with photoelectric astrolabe of the Yunnan Observatory from 2010.0 to 2016.05. Marks show large earthquakes: (1) 2013-04-20, Lusha, M=7.0; (2) 2014-05-30, Yingjiang, M=6.1; (3) 2014-08-03, Ludian, M=6.5; (4) 2014-10-07, Jinggu, M=6.6.

Interconnection of the Earth rotation and its magnetic field has been actively investigated during past decades. Paper Modiri et al. [2021] is devoted to investigation of the investigation of the connection of the celestial pole motion (CPM) and variations of the geomagnetic field (GMF). This study used the celestial pole offsets (CPO) time series obtained from very long baseline interferometry (VLBI) observations and data such as spherical harmonic coefficients, geomagnetic jerk, and magnetic field dipole moment from a state-of-the-art geomagnetic field model to explore the correlation between them. In this study, we use wavelet coherence analysis to compute the correspondence between the two non-stationary time series in the time–frequency domain. Our preliminary results reveal interesting common features in the CPM and GMF variations, which show the potential to improve the understanding of the GMF contribution to the Earth's rotation. Special attention was given to correspondence between the free core nutation (FCN) and GMF (Fig. 5) and potential time lags between geomagnetic jerks and rotational variations.

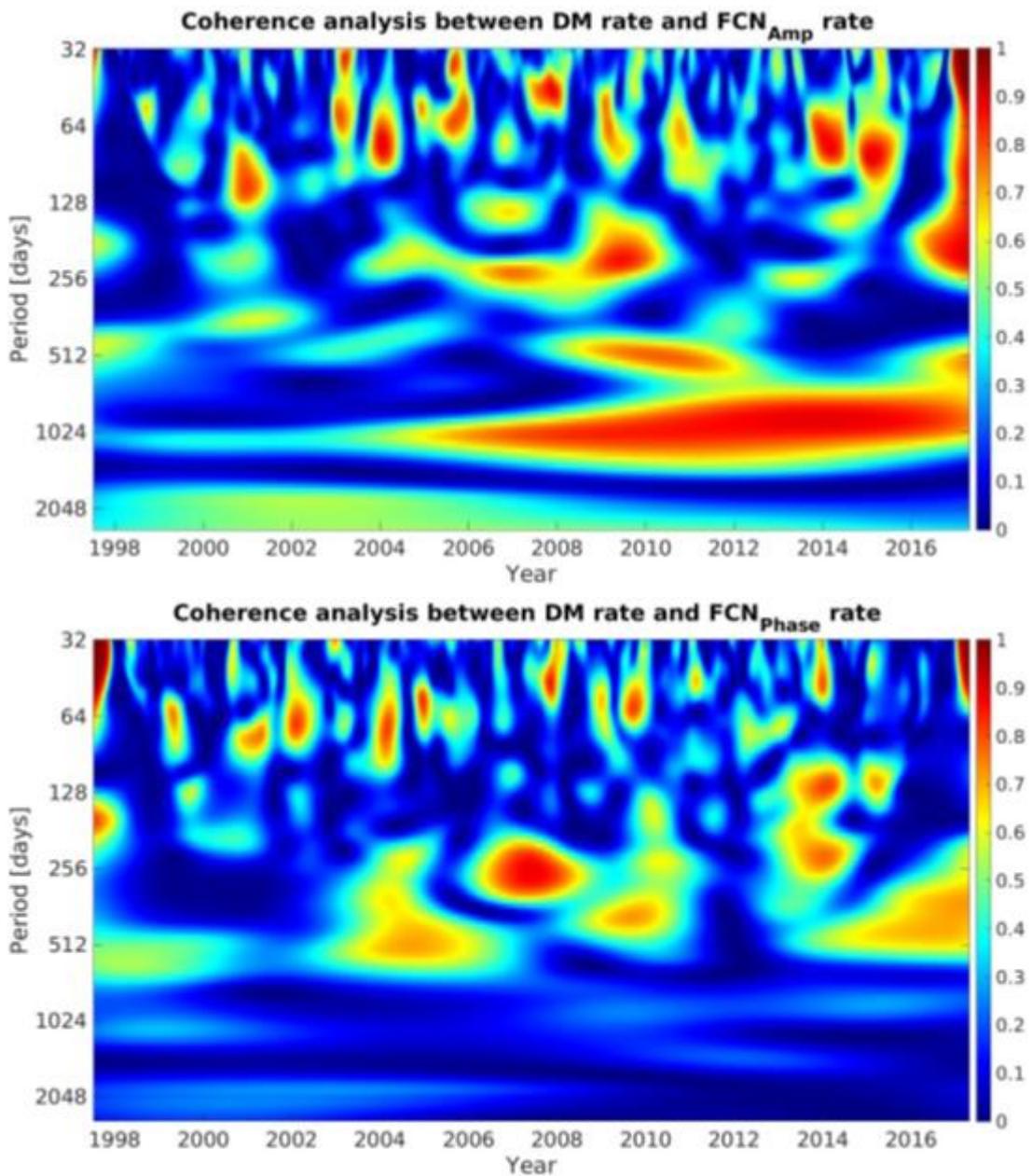

Fig. 5. Wavelet coherence between the FCN and GMF dipole moment.

Since 2016 the Earth rotation has been accelerating and LOD reached its minimum in 2022, which attracted much attention of world media [Zotov, Bizuar, 2021; Carter, 2022]. Chandler wobble of the pole has almost disappeared in 2017–2020 and its phase seems to be changing as in 1930-s [Zotov et al., 2022a]. The monography by [Zotov, 2022b] is devoted to the consideration of geophysical processes that cause the motion of the pole and changes in the length of the day.

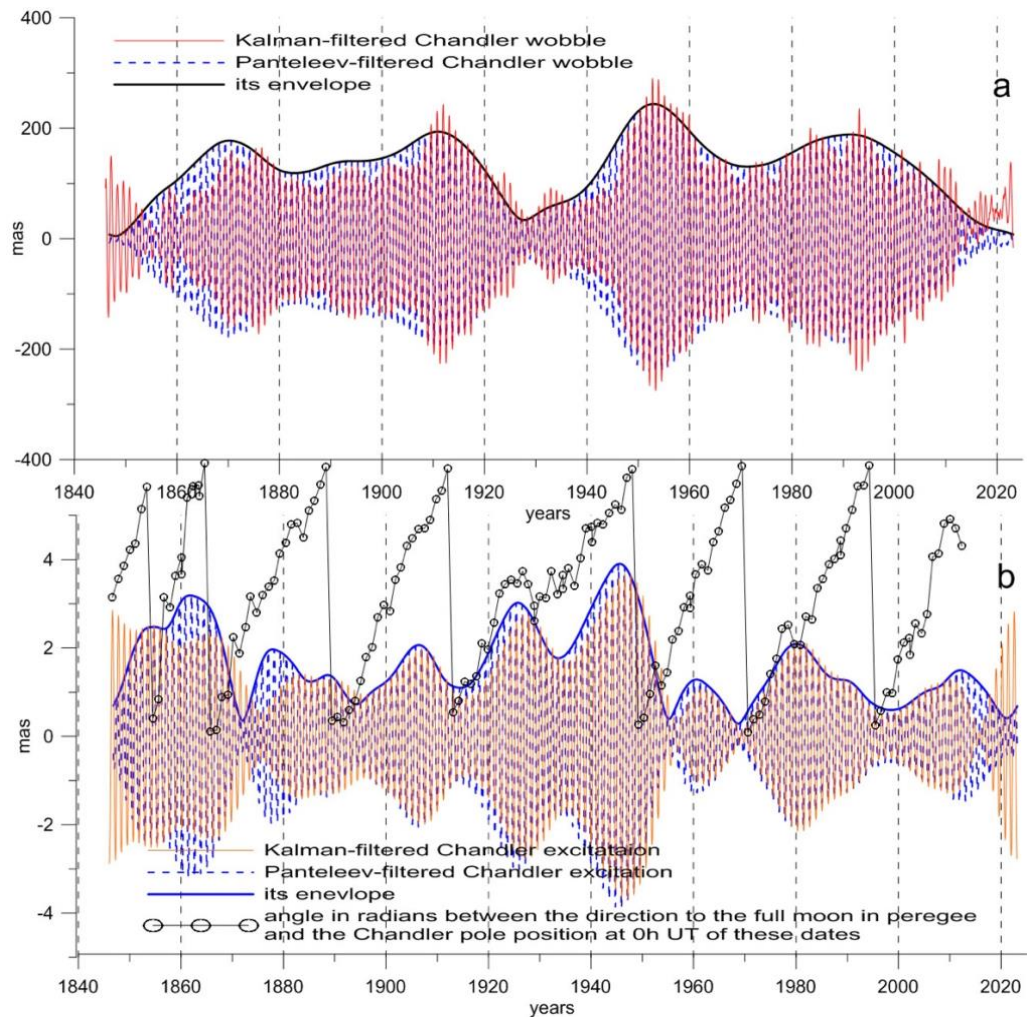

Fig. 6. (a) Chandler wobble extracted by Kalman filtering and by Panteleev filtering in frequency domain. (b) Chandler excitation in comparison with the angle between the perigee of the lunar orbit and Earth's Chandler pole position for syzygies in perigee days at $0^h$UT.

Particular attention is paid to the Chandler wobble. The method of its extraction and geodesic excitation reconstruction is developed. Atmospheric and oceanic excitation functions are investigated in the Chandler frequency band. The book also covers the Earth's gravity field from GRACE satellites, its decomposition and analysis, theory of the Earth rotation, Euler-Liouville equations generalization. The relationships of the Earth's rotation and climate processes are discussed. Earth rotation parameters predictions are continued to be performed at SAI MSU on daily basis. In the light of newly found 6-year oscillations in the Earth systems [Pfeffer et al., 2023], in particular in LOD [Zotov et al., 2020], old ideas of N. Sidorenkov about perigee and node of the lunar orbit revolution come back to life. Recent studies by L. Zotov have revealed, see Fig 6, that ~22-year cycle of the angle between the Chandler pole and the direction to the perigee can lie in the origin of Chandler wobble modulations, Markowitz wobble [Zotov et al., 2022c] and LOD decadal variability [Zotov et al., 2023].

# Positioning and Applications

Gorshkov V.[1], Gusev I.[2], Malkin Z.[1], Shestakov N.[3]

[1]Pulkovo Observatory, Saint Petersburg, Russia

[2]Joint Stock Company «Central Research Institute for Machine Building», Moscow Region, Russia

[3]Far Eastern Federal University, Vladivostok, Russia

### *Methods and techniques*

New methods for improving the accuracy of coordinate-time measurements are considered in [Bezmenov et al., 2022]. The GGOS Global Geodetic Observation System proclaimed the goal of achieving millimeter accuracy in determining the coordinates of reference points. To achieve such accuracy, it is necessary to equip measurement points with new generation VLBI, GNSS and SLR. However, since the accuracy targets were at the boundary of the capabilities of modern measuring instruments, it was also necessary to improve models and methods for both secondary processing of measurements and their pre-processing.

During the preliminary processing of measurements, the task of rejecting outliers - the results of rough measurements - arises. This problem is closely related to the problem of finding an unknown trend (usually polynomial) in measurement data. Obviously, an incorrect trend definition can lead to incorrect outlier detection, and good measurements can be rejected and inaccurate ones left, which ultimately negatively affects the accuracy of the final result. An-other problem is the detection of coarse measurements in data with removed trend. One object in creating detection algorithms in this case is to minimize the amount of rejected data. Moreover, in the case of GNSS measurements, the loss of some part of the data at the preliminary processing stage may not have a significant effect on the final result due to the huge number of navigation measurements. At the same time, when processing satellite laser rangefinder (SLR) measurements, each individual measurement is important.

To solve the problem of automatic detection of rough measurements at the stage of their preliminary processing work, algorithms developed by I.V. Bezmenov, S.L. Pasynok, et al., were involved. Initially, they were designed to reject coarse GNSS measurements and showed their effectiveness compared to other known methods (fewer computational operations, fewer rejected points with the same rejection threshold). As a result of using these algorithms for the preliminary processing of laser measurements, I.Yu. Ignatenko managed to achieve greater accuracy of formed normal points with a smaller number of rejected data. The effectiveness of the proposed algorithm in pre-processing SLR measurements under conditions of abnormal atmospheric refraction was also shown.

The algorithm for determining jumps in the Melbourne-Wubben combination formed from code and phase measurements in global navigation satellite systems is presented in [Bezmenov et al., 2019].

The paper [Kaftan et al., 2022] presents the methodology and results of assessing the accuracy of GNSS measurements on a reference geodetic basis.

The velocity vector field of the dense network of regional GNSS reference stations provides the basic data for the analysis of deformation processes in the faults, for study of the intraplate structure of region and other geodynamical studies. For research of these problems the database of GNSS station velocities (VDB) approximately covering the East European Craton was created and supported now with colleagues of various geodetic organizations [Gorshkov, Shcherbakova, 2019]. For the beginning of 2019 the VDB has uniformly processed data for more than 350 GNSS stations. The VDB is available on the website of the Pulkovo observatory, where the full description of used technique is also provided:

http://www.gaoran.ru/russian/database/station/databasev_eng.html

The initial GNSS observations are processed by the Gipsy 6.4 (JPL) software with the use of PPP strategy. All standard model parameters and corrections are taken into account (absolute antenna calibration, final orbital parameters and clock corrections (IGS14), Earth orientation parameters IERS (C04), tropospheric model VMF1/ECMWF, all solid Earth's tides, including corresponding polar tides, ocean tidal loads (GOT4.8) and ionospheric effects, including second-order terms in the model IONEX). All mass loading corrections also taken into account according to IMSL (http://massloading.net/). Seasonal variations in the GNSS position series are estimated and then removed.

The mean errors of all velocity components of GNSS station velocities are given in the VDB as for for the normal as for flicker noise distribution.

The paper [Gorshkov et al., 2021] contains description and principles underlying the development of the velocities database of almost 500 GNSS stations with an observational duration of more than two years. The stations are mainly located on the territory of Eastern and Northern Europe. Various station motion models, a multi-step technique for filtering of outliers, estimation of errors for station velocities on the assumption of different types of noise distribution as well as accounting discontinues in position time-series of stations are used when creating this base. More than 60% of the stations are owned and constantly updated with data from various geodetic institutions of the Russian Federation. All data of the stations is presented for the first time within one homogeneously processed database. Velocity field based on stations in the database used for the study of the motion of the East European Craton and its parts – Baltic Shield and Russian Plate. An example of atmospheric monitoring for a particularly dense network of this base around the Gulf of Finland is also given.

The paper [Maciuk et al., 2021] devoted to testing the product quality of Galileo and GPS on-board oscillators. The final clock product provided by the IGS (International GNSS Service) with a 30/300 s sampling interval for satellites and stations respectively is characterized by accuracy of $75 \pm 20$ ps RMS (Root Mean Square). Since mid-2012, the MGEX (Multi-GNSS Experiment) experiment provides full products containing all available GNSS (Global Navigation Satellite Systems) satellite signals. In this paper, the phase, frequency and stability of Galileo and GPS (Global Positioning System) on-board oscillators based on the MGEX data are analysed. Moreover, the authors also determined the noise type for specific on-board oscillators depending on GNSS type and frequency standard. The analysis showed different characteristics of clock correction phase and frequency for both satellite systems. Also, each system and clock type has different noise for each averaging time. In order to distinguish more types of noises, three different variants of stability analysis were conducted: Allan deviation (ADEV), modified Allan deviation (MDEV), and Hadamard deviation (HDEV).

### *Troposphere sounding*

GNSS methods are widely used to study integrated water vapor (IWV) changes in the Earth's surface layer of the atmosphere and their relationship with various processes - typhoons and cyclones, dust storms, forest fires, technogenic atmospheric pollution, and others. In [Shestakov et al., 2021], based on GNSS observations at two permanent stations located in the continental and coastal parts of Primorsky Krai (Russian Far East), the change in IWV in the continent-ocean transition zone is studied. Using the measurement information at the nearest stations of the IGS global GNSS network and upper-air sounding data, the high accuracy and reliability of the atmospheric moisture content estimates obtained were confirmed. Changes in IWV for the period 2015–2019 were studied at the measurement points, empirical approximation models of annual IWV variations were built, and the obtained estimates were compared with the data of the global GFS and Reanalysis ERA5 models. The daily changes in the concentration of water vapor in the atmosphere, as well as its change during the passage of typhoons, accompanied by massive precipitation, have been studied. It has been established that more than 60% of massive precipitation (>20 mm) falls within 3–9 hours at the IWV decline after a sharp increase in the integrated moisture content recorded by GNSS methods. The high accuracy and frequency of determining IWV (up to 1 Hz), together with the high speed of obtaining information about the change in IWV from GNSS observations, open up broad prospects for the use of GNSS meteorology in the forecasting practice of hydrometeorological services in the Russian Federation. The paper [Kishkin et al., 2022] presents an analysis of estimates of the integral content of water vapor in the atmosphere based on continuous GPS/GLONASS observations for the period 2017–2019. at thirteen points of the GNSS network in Primorsky Krai. The obtained IWV estimates are compared with the data of the Global Forecast System. For comparison, 16 nodes of the computational grid of the model closest to each sensor were selected. It is shown that the correlation of the IWV GNSS estimates with the GFS data at the time of the forecast release is on average >0.90; starting from the forecast lead time of 48 hours, the correlation coefficient decreases to 0.60. Correlation with GFS data in the warm season is 0.85–0.97, in the cold season ≤0.60. An analysis of the spatial distribution of the correlation coefficients showed that the measured IWV values are linearly related to the model PWEA values, which refer to grid nodes that have a smaller height difference with the GNSS point. It is concluded that it is promising to "assimilate" the results of GNSS sounding in regional atmospheric models.

The dynamics of the atmospheric integrated precipitable water vapor (IPWV) over the Leningrad and adjacent regions has been investigated according to the GNSS database maintained at the Pulkovo Observatory [Gorshkov et al., 2019]. The data of 65 GNSS-stations was used for this study. The required for IPWV estimation the mean atmospheric temperature was calculated by using as surface temperature from nearest meteorological stations as interpolated from the global atmospheric database NCEP/NCAR Reanalysis-1. The comparisons of GNSS IPW estimations with based on surrounding the region radiosondes and water vapor radiometer in Svetloe station have been made. All results are in good agreement. The temporal dynamics of the IPW spatial distribution was made for study region. The trends of IPWV ($0.34 \pm 0.006$ mm/10year) have been estimated according 31 longest GNSS-station series in this area.

# Common and Related Problems

Kaftan V.[1], Malkin Z.[2], Shestakov N.[3]

[1]Geophysical Center of the Russian Academy of Sciences, Moscow, Russia
[2]Pulkovo Observatory, Saint Petersburg, Russia
[3]Far Eastern Federal University, Vladivostok, Russia

The study [Mazurova, Petrov, 2022a, b] are devoted to the Sagnac effect, one of the main relativistic effects, which must be taken into account for synchronization of high-precision frequency standards. The theoretical substantiation of the Sagnac effect is presented both within the framework of the kinematic effect of the special theory of relativity and within the framework of the general theory of relativity, where the effect is analyzed as a result of the action of the potential of centrifugal forces. For each of the theories, two different approaches are used, which allows one to delve deeper into the physical meaning of the phenomenon. A description of the effect in real conditions on the surface of a rotating Earth, when optical fiber is used to synchronize high-precision frequency standards, is presented.

In [Shestakov et al., 2021; Perevalova et al., 2021], the effects of lithospheric-ionospheric interactions initiated by the Sarychev Peak volcano eruption and an underground nuclear test conducted on September 3, 2017 in North Korea are considered. The article [Shestakov et al., 2021] studied in detail the parameters of ionospheric anomalies caused by a series of strong explosions on June 14-15, 2009. The amplitude-frequency characteristics and features of the propagation of such anomalies at distances up to 900 kilometers or more have been determined. Authors also obtained data on a possible off-center position of the secondary source of ionospheric anomalies relative to the volcano.

In [Perevalova et al., 2021], based on the analysis of data from several networks of stations receiving signals from the GPS and GLONASS systems, studies of ionospheric disturbances initiated by the North Korean underground nuclear test (explosion) on September 3, 2017 were carried out. Disturbances in the ionosphere were observed on a large number of receiver - satellite beams. The shape of the disturbances generated by the underground nuclear test was noticeably different from the shape of the ionospheric anomalies observed after earthquakes. Ionospheric disturbances began to be recorded ~8 min after the explosion and were observed for more than 5 h after it. It is shown that within 1.5 hours after the explosion, predominantly traveling ionospheric disturbances (TIDs) were recorded with periods from 1.0 to 9.5 minutes, propagating from the epicenter with average velocities of the order of 580, 250, and 130 m/s. These TIDs can be attributed to acoustic waves induced in the atmosphere by an underground nuclear test. Approximately 60 min after the explosion, a long-lived (observed for more than 3.5 h) region of low-moving disturbances of the ionospheric plasma began to form above the epicenter, the velocity of which was ~8 m/s. The nature and formation mechanisms of this area require further research and modeling.

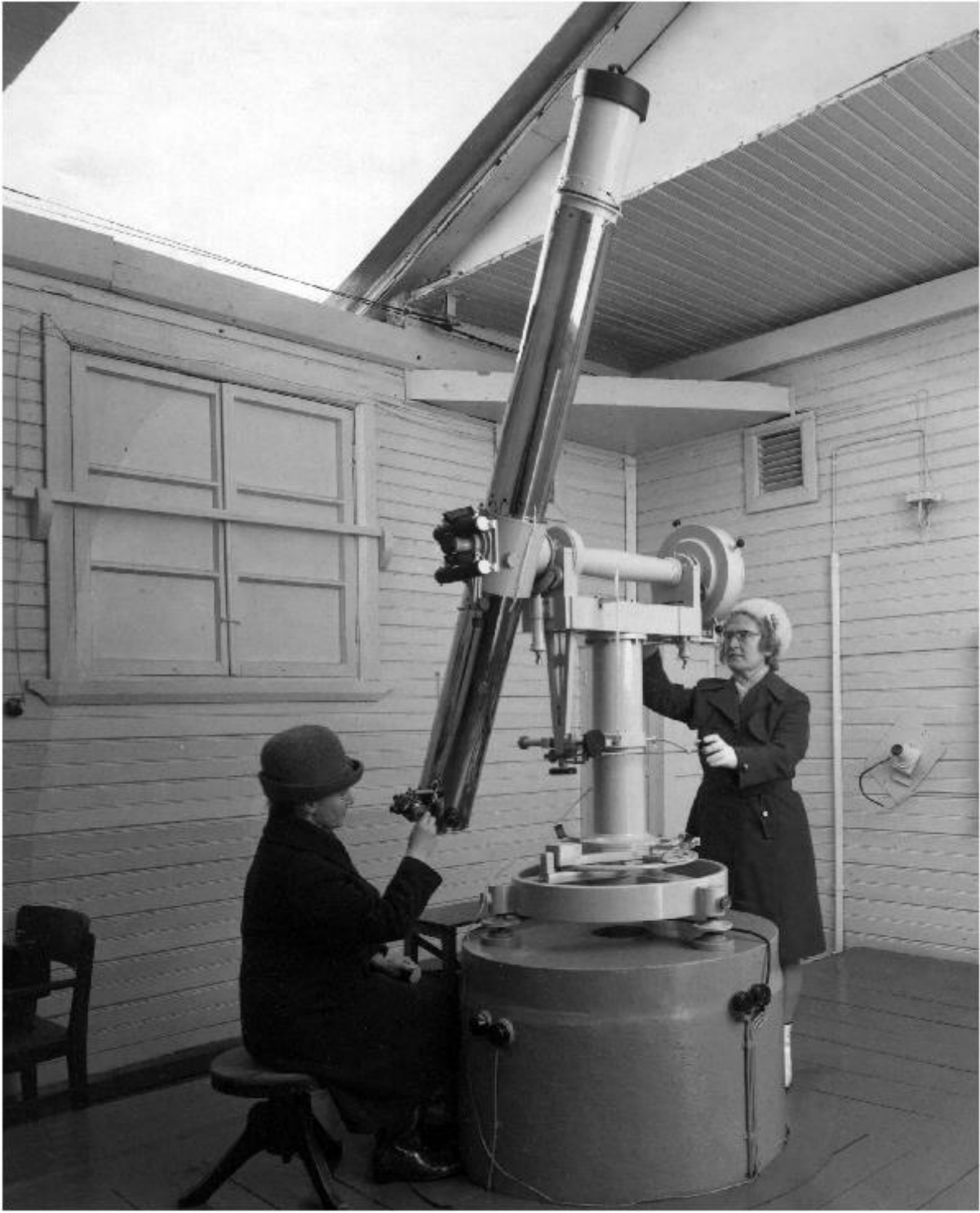

Fig. 1. L. D. Kostina (left) and N. R. Persiyaninova (right) at the Pulkovo Observatory at the telescope ZTF-135.

In publications [Komitov, Kaftan, 2019; 2021; Komitov, Kaftan, 2020] presents the results of a statistical analysis of the relationship between solar activity and climate and terrestrial tectonic processes.

The identification of tectonic linear structures by means of discrete mathematical analysis was proposed in [Agayan et al., 2021].

The strategy for searching for outliers in series of noisy data with an unknown trend is considered in [Bezmenov et al., 2022].

The work [Brovar et al., 2022] analyzes the problems of the development of geodesy in Russia on the example of one of the doctoral dissertations.

Information about holding the International Conference JOURNEES 2019 in Russia is presented in the article [Pasynok, 2029].

Paper Malkin et al. [2020] is devoted to the memory of two outstanding Pulkovo latitude observers Lidia Kostina and Natalia Persiyaninova. Lidia Dmitrievna Kostina and Natalia Romanovna Persiyaninova left a bright mark in the history of the Pulkovo Observatory, as well as in the history of the domestic and international latitude services. In the first place, they were absolute leaders in the latitude observations with the famous zenith telescope ZTF-135. In 1954-2001, they obtained together more than 66'000 highly accurate latitudes, which make about 2/3 of all the observations made by 23 observers with the ZTF-135 after the WW2. They also provided a large contribution to investigation of instrumental errors, methods of the data analysis, developing of the observing programs. Their results in studies of latitude variations and polar motion were also highly recognized by the community.